\documentclass[twocolumn,amssymb,aps,prc,superscriptaddress]{revtex4-2}
\usepackage{graphicx}
\usepackage{physics}
\usepackage{color}
\usepackage[colorlinks=true,linkcolor=blue,citecolor=blue,urlcolor=blue]{hyperref}

\newcommand{\nnb}{\nonumber}
\newcommand{\1}{^\prime}
\newcommand{\pythia}{{\sc pythia}}

\begin{document}


\title{ Study of the $\eta$ to $\pi^0$ Ratio in Heavy-Ion Collisions }

\author{Yuanjie \surname{Ren}}%
\email[]{yuanjie@mit.edu}
\affiliation{Department of Physics, Massachusetts Institute of Technology, Cambridge, Massachusetts 02139, USA}
\author{Axel \surname{Drees}}
\email[]{axel.drees@stonybrook.edu}
\affiliation{
Department of Physics and Astronomy, Stony Brook University, Stony Brook, New York 11790, USA}

\date{\today}

\begin{abstract}
We demonstrate that the $p_T$ dependence of the $\eta/\pi^0$ ratio is universal within a few percent for high energy $p$+$p$,  $p$+A and $d$+A collisions, over a broad range of collision energies. The $\eta/\pi^0$ ratio increases with $p_T$ up to 4 to 5 GeV/$c$ where it saturates at a nearly constant value of 0.487$\pm$0.024. Above $p_T = 5$ GeV/$c$ the same constant value is also observed in A+A collisions independent of collision system, energy, and centrality. At lower $p_T$, where accurate $\eta/\pi^0$ data is absent for A+A collisions, we estimate possible deviations from the universal behavior, which could arise due to the rapid radial hydrodynamic expansion of the A+A collision system. For A+A collisions at RHIC we find that possible deviations are limited to the $p_T$ range from 0.4 to 3 GeV/$c$, and remain less than 20\% for the most central collisions. 

\end{abstract}

\maketitle
\section{\label{Sec:Intro}Introduction}

Photons are generally considered ideal probes to study the quark gluon plasma (QGP) created in heavy ion collisions \cite{Shuryak}, since they have a long  mean free path and leave the collision volume without final state interactions. Of particular interest are low momentum or thermal photons with energies of up to several times the temperature of the QGP. The measurement of thermal photons has only recently been possible with the advance of the heavy ion programs at RHIC \cite{Adare:2008ab, Adare:2014fwh,Adare:2018wgc} and LHC \cite{Adam:2015lda}. 

One of the  experimental key challenges for these measurements is to estimate and subtract photons from hadron decays that constitute the bulk of photons measured in experiments. The two major contributions of photons result from $\pi^0\rightarrow \gamma+\gamma$ and $\eta\rightarrow \gamma+\gamma$ decays. Precise knowledge of the parent $\pi^0$ and $\eta$ $p_T$ spectra is necessary to estimate the decay photon background. While spectra of pions from heavy ion collisions are well measured at RHIC and LHC, less data exists for $\eta$ spectra, in particular below $p_T$ of 2 GeV/$c$. Therefor experiments need to make assumptions how to model the $\eta$ spectra below 2 GeV/$c$, which leads to sizable systematic uncertainties.  Frequently, experiments have based this extrapolation on the hypothesis of transverse mass $m_T$ scaling of meson spectra \cite{Adare:2014fwh,Adare:2018wgc,Adam:2015lda}. However, it is known since the late 1990's \cite{Agakishiev:1998mw} and was recently pointed out again \cite{Altenkamper:2017qot} that $m_T$ scaling does not hold below 3 GeV for the $\eta$ meson. 

In this paper we propose a new empirical approach to model the $\eta$ spectrum that is based on the universality of the $\eta/\pi^0$ ratio across collision systems, beam energies, and centrality selections in heavy ion collisions. With a good understanding of the $\eta/\pi^0$ ratio as function of transverse momentum $p_T$ and measured $\pi^0$ spectra, which are readily available for many collision systems, one can construct a more accurate $p_T$ distribution for $\eta$ mesons.   

The paper is organized as follows. In the next section we elaborate  more on the failure of $m_T$ scaling. In section III we will discuss two empirical fits and a Gaussian Process Regression (GPR) to describe the $\eta/\pi^0$ ratio for $p$+$p$ and $p$+A collisions, and document in section IV the universality of $\eta/\pi^0$ across  different collision systems ($p$+$p$, $p$+A, A+A), energies, and collision centrality. In Section V, we estimate possible deviation from the universal trend at low $p_T$ due to radial flow in heavy ion collisions. We provide our result for $\eta/\pi^0$ for RHIC and LHC energies with systematic uncertainties in the final part.   


\section{The Failure of Transverse Mass Scaling}

For measurements of direct photons from heavy ion collisions, the photons from $\eta$ and heavier meson decays are frequently estimated using measured $\pi^0$ spectra in conjunction with the  $m_T$ scaling hypothesis. A typical implementation of this method \cite{ppg088} starts with a fit to the $\pi^0$ spectra with a functional form like a modified Hagedorn function \cite{Hagedorn:1965st}: 
\begin{align}
\frac{1}{2\pi p_T}\frac{\dd^2N}{\dd y\dd p_T}=A(M_X)\qty(e^{-ag(p_T)-b g(p_T)^2}+\frac{g(p_T,M_X)}{p_0})^{-n}\nonumber\\
\label{Eq:modified_hagedorn}
\end{align}
with $M_X$ being the meson mass and  $g(p_T,M_X)=\sqrt{p_T^2+m_X^2-m^2_\pi}$. In this implementation the spectra of the $\eta$ and heavier mass mesons follow the same distribution with respect to transverse mass $m_T\equiv \sqrt{m^2+p_T^2}$ as the $\pi^0$. The normalisation constant $A(M_X)$ is the only free parameter, all other parameters are fixed by the fit to the $\pi^0$ data. $A(M_X)$ is fitted to experimental data whenever such data exists. 

\begin{figure}
\includegraphics[width=3.0in]{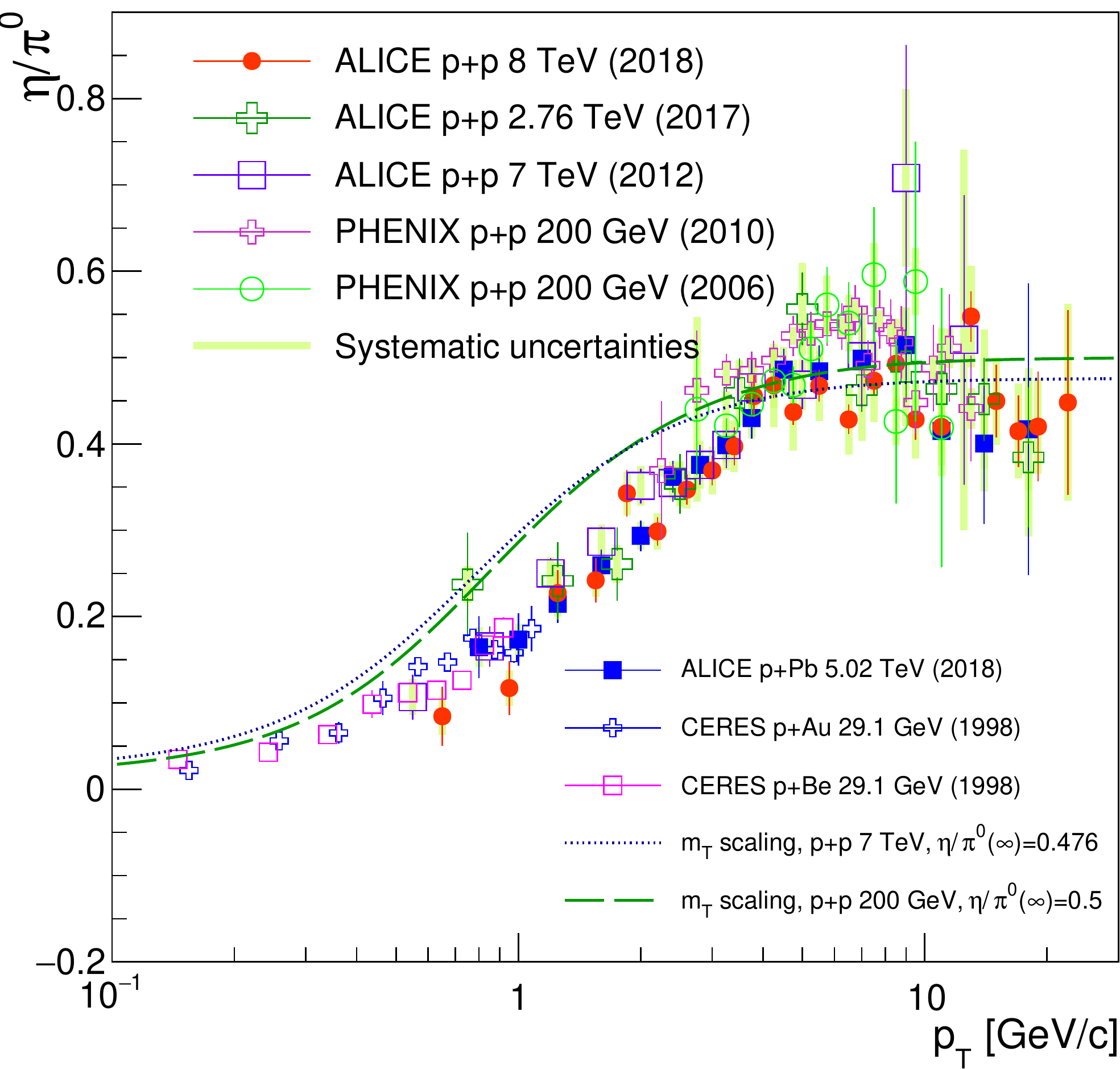}
\caption{The $\eta/\pi^0$ ratio of p+p and p+A collisions. Also plotted are the $\eta/\pi^0$ determined by  $m_T$-scaling and from a \pythia \ calculation.}\label{Fig:etapi_mT_scaling}
\end{figure}

Fig.~\ref{Fig:etapi_mT_scaling} compiles available data of $\eta/\pi^0$ for $p$+$p$ \cite{Acharya:2017tlv,Abelev:2012cn,Adler:2006bv,Acharya:2017hyu,Adare:2010cy} and $p$+A   \cite{Agakishiev:1998mw,Acharya:2018hzf} collisions. Also shown on the figure is the result of $m_T$ scaling for two different normalisation constants  $A(M_\eta)$ \cite{Adler:2006bv,Abelev:2012cn,Adler:2003pb} and the expectation from a \pythia-6 calculation from \cite{Sjostrand:2000wi, Adler:2006bv}. While \pythia \ and the $m_T$ scaling hypothesis agree well, a significant deviation from the data is seen at low $p_T$. This was originally discovered at the CERN SPS by CERES/TAPS \cite{Agakishiev:1998mw} more than 20 years ago and recently confirmed by ALICE at the LHC \cite{Acharya:2017tlv}. Clearly the $m_T$ scaling hypothesis is not correct and should not be used to extrapolate meson spectra to low $p_T$ for systems where no data exists.

\section{Description of the $\eta/\pi^0$ ratio for p+p and p+A collisions}\label{Sec:empirical_etapi}

The quantitative agreement of the $\eta/\pi^0$ data shown in Fig.~\ref{Fig:etapi_mT_scaling} is striking, consider the data covers more than 2 orders of magnitude in collision energy. 
In this section we will test different methods to obtain an  
empirical description of $\eta/\pi^0$. The first two methods (A,B) fit a functional shape of the ratio, while the third method (GPR) is a Gaussian Process Regression that does not assume a specific functional shape. All methods yield similar results below 10 GeV/$c$, at larger $p_T$ the deviations are sizable and we will include these deviations in our evaluation of systematic uncertainties. 

\subsection{Empirical fit A}
Method A starts with a ratio of two functions of the form given in Equation~\ref{Eq:modified_hagedorn}. The $m_T$-scaling hypothesis is used to reduce the number of parameters: 

\begin{equation}\label{Eq:empirical_fit_A}
R^{\eta/\pi^0}(p_T)=R^\infty \frac{
   \qty( e^{-a\cdot g(p_T)-b\cdot g(p_T)^2}+ \frac{g(p_T)}{p_0}  )^{-n} }
{ \qty( e^{-a p_T-bp_T^2}+\frac{p_T}{p_0}  )^{-n} }.
\end{equation}

The advantage of this method is that it preserves a realistic functional form for the $p_T$ spectra with an exponential decrease at low $p_T$ and power law shape at high $p_T$. In principle, this ensures that at high $p_T$ the $\eta/\pi^0$ ratio approaches a constant value $R^\infty$. However, unlike starting from the $\pi^0$ spectrum, the parameters are fitted to the $\eta/\pi^0$ ratio from p+p and p+A collisions shown in Fig.~\ref{Fig:etapi_mT_scaling}. We achieve a good fit, though the values of the fit parameters are nonphysical and do not describe the individual $p_T$ spectra. The result is depicted in Fig.~\ref{Fig:pp_pA_empirical}.  

The band represents the total uncertainty of the fit function from two sources, the uncertainty of fit parameters, and the systematic uncertainties from data points.  The former can be calculated analytically thanks to the explicit fit function 
while the latter can be obtained via a ``data shuffling approach" which uses a Monte Carlo technique to vary individual data sets within their systematic uncertainties. This approach is discussed in Appendix~\ref{Appendix:shuffle}. The total uncertainty shown on the figure represents the quadratic sum of statistical and systematic uncertainties.

\subsection{Empirical fit B}
The second empirical fit function has a very similar form, except that normalization of the exponential and power law component in the numerator are decoupled by introducing an additional parameter.
This is implemented such that $R^\infty$ remains the asymptotic value at high $p_T$.   

\begin{equation}\label{Eq:empirical_fit_B}
R^{\eta/\pi^0}(p_T)=A\frac{
   \qty( e^{-a\cdot g(p_T)-b\cdot g(p_T)^2}+\qty(\frac{R^\infty}{A})^{-\frac{1}{n}} \frac{g(p_T)}{p_0}  )^{-n} }
{ \qty( e^{-a p_T-bp_T^2}+\frac{p_T}{p_0}  )^{-n} }.
\end{equation}

The handling of fit and the calculation of the uncertainties is identical to Method A. The result is also shown in Fig.~\ref{Fig:pp_pA_empirical}. In contrast to Method A, which only gradually approaches the asymptotic value at high $p_T$, Method B reaches the constant at $p_T$ of about 5 GeV/$c$ and at a lower $R^\infty = 0.487\pm0.024$ value, which will be used as a reference throughout this article. We note that the change to the constant value is rather abrupt.

\subsection{Gaussian Process Regression (GPR)}
Both previous methods have a built-in assumption that the $\eta/\pi^0$ has a constant asymptotic value at high $p_T$. However, the data suggest that there might be a maximum around 8 GeV followed by a decrease towards higher $p_T$. In order to avoid any assumptions about the shape we resort to a machine learning technique called Gaussian  Process Regression (GPR), which possesses no physical knowledge but gives full trust to the data it is given. Details about the GPR can be found in \cite{MIT_GPR}, and comments about the specific implementation we use are summarised in Appendix \ref{Appendix:GPR}. In general the GPR works best in the region where many consistent data points are available. Less data points or inconsistent data sets lead to larger uncertainties, and unlike the fitting methods the GPR can not reliably extrapolate much beyond the range covered by data.

\begin{figure}
\includegraphics[width=2.8in]{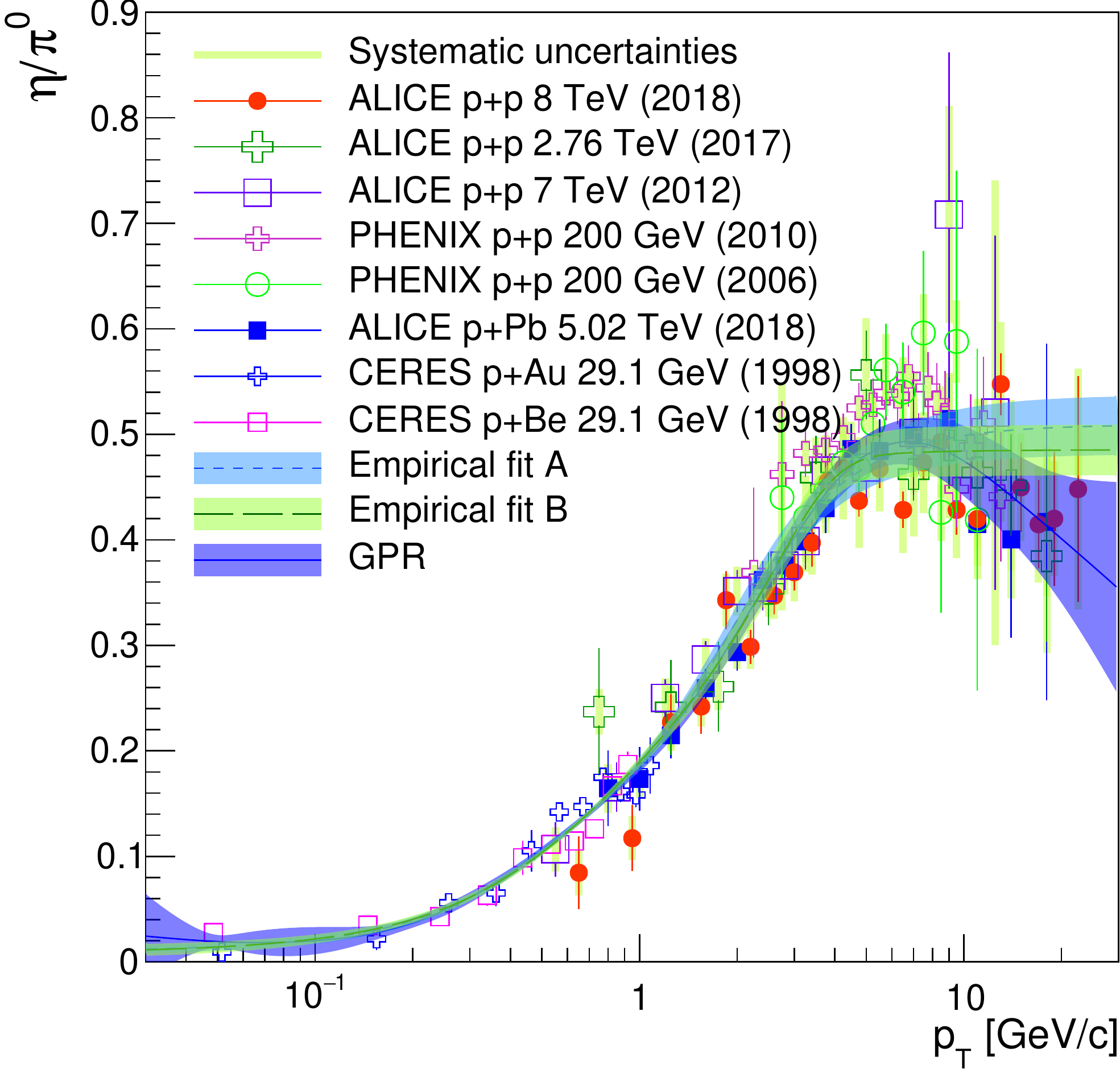}\ \ \ \ \ 
\caption{Data for the $\eta/\pi^0$ ratio from $p$+$p$ and $p$+A collisions compared to three different methods to describe the data with a universal shape: empirical fit A, empirical fit B, and GPR. }\label{Fig:pp_pA_empirical}
\end{figure}

The result of the GPR is presented in  Fig.~\ref{Fig:pp_pA_empirical}, with the band indicating the uncertainties. Over most of the $p_T$ range the GPR gives an equally good description of the data compared to Methods A and B. As expected, it follows the data and peaks near 8 GeV/$c$. Towards higher $p_T$ $\eta/\pi^0$ from the GPR decreases. Whether the drop at high $p_T$ is physical or an artefact of different data sets with different $p_T$ ranges not being perfectly consistent in the range from 3 to 10 GeV/$c$ will only be resolved with more precise data. 

\begin{figure}
 \includegraphics[width=2.8in]{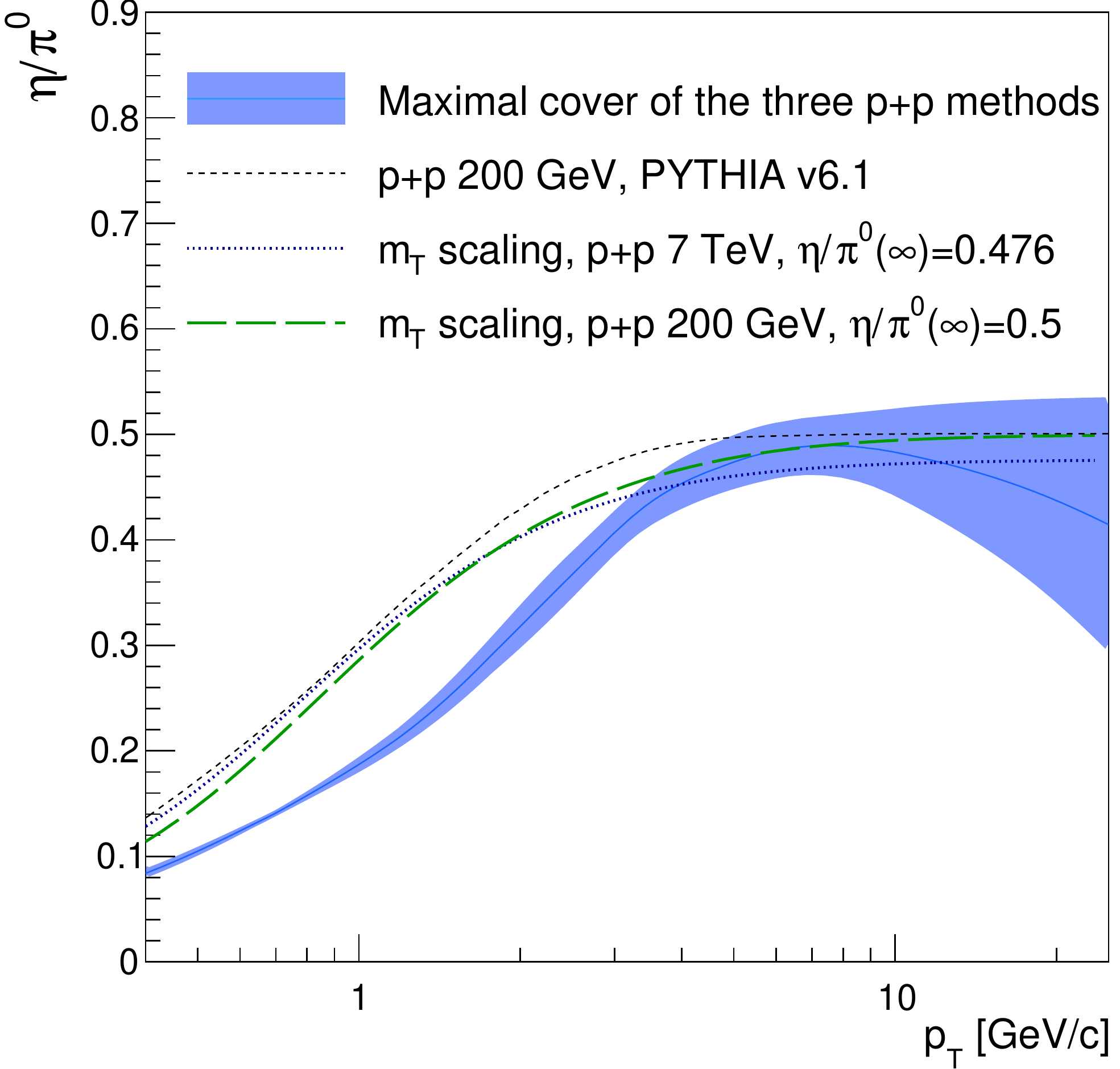}
\caption{Result of combining the three empirical methods to one universal estimate of $\eta/\pi^0$ as function of $p_T$. Also shown for reference are the estimates based on the $m_T$ scaling hypothesis and the result of a \pythia \ calculation, both from Fig.~\ref{Fig:etapi_mT_scaling}. }
\label{Fig:etapi_universal}
\end{figure}

Since we do not know the correct functional form of $\eta/\pi^0$, in particular at high $p_T$, we combine the results obtained with the three methods as our best estimate for a universal $\eta/\pi^0$ ratio for $p$+$p$ and $p$+A collisions. This is achieved by assigning every $p_T$ value the minimum of the lower uncertainty range of the three methods as the lower bound and the maximum as the upper bound. The average of the lower and upper bound is used as central value. In the following we will use $(\eta/\pi^0)_{pp}^{mc}$ to refer to this combined result, with the superscript $mc$ referring to maximal coverage of uncertainties. The result is given in  Fig.~\ref{Fig:etapi_universal} and compared to the $m_T$-scaling prediction as well as the \pythia\   calculation already shown in Fig.~\ref{Fig:etapi_mT_scaling}. One can see that all of the theoretical predictions overestimate the ratio for $p_T$ below 3-4 GeV/$c$.

\section{Universality of $\eta/\pi^0$ ratio systems at high $p_T$}\label{Sec:universal_etapi}

In the previous section we established that the $\eta/\pi^0$ ratios measured in $p$+$p$ and $p$+A collisions are consistent with being constant at high $p_T$ with a value of $R^\infty = 0.487\pm0.024$ (Section III.B). Here we demonstrate that all available data from $p$+$p$, $p$+A, and A+B collisions listed in Table~\ref{Tab:Data} are consistent with this $R^\infty$ value independent of the collision energy, collision system, or collisions centrality. 
\begin{table}
      \caption{References and systems quoted in this article are collected in this table. For each A+A system, if different centralities have different $p_T$ ranges, the one of the minimum bias is presented.\label{Tab:Data}}
      \ruledtabular
      \begin{tabular}{c| c| c| c| c}
         System     &   Experiment      &   $\sqrt{s_{_{NN}}}$& $p_T$ range [GeV/c] & Ref.   \\ \hline 
          p+p    &   CERN WA70    &   23 GeV& $4-6$ &\cite{ref80}   \\ \hline
          p+p    &   Fermilab E706    &   31.6 GeV& $3.25-7.5$ & \cite{Apanasevich:2002wt}  \\ \hline
          p+p    &   Fermilab E706    &   38.8 GeV& $3.25-9$ & \cite{Apanasevich:2002wt}  \\ \hline
          p+p    &   PHENIX    &   200 GeV& 2.75-11 &  \cite{Adler:2006bv}  \\ \hline  
          p+p    &   PHENIX    &   200 GeV& 2.25-13 & \cite{Adare:2010cy}  \\ \hline 
          p+p    &   ALICE   &   2.76 TeV & 0.75-18&  \cite{Acharya:2017hyu} \\ \hline 
           p+p    &   ALICE    &   7 TeV & 0.55-12.5 &   \cite{Abelev:2012cn} \\ \hline 
          p+p    &   ALICE    &   8 TeV & 0.65-22.5 & \cite{Acharya:2017tlv}  \\ \hline
           p+Au&  CERES-TAPS    & 29.1  GeV &  $0.05-1.1$   &\cite{Agakishiev:1998mw} \\ \hline 
          p+Be &CERES-TAPS  & 29.1 GeV & $0.05-1$   &\cite{Agakishiev:1998mw}\\  \hline
          p+Pb &ALICE   &5.02 TeV & $0.8-18$ &\cite{Acharya:2018hzf} \\ \hline
          Cu+Au & PHENIX &200 GeV & $2.25-19 $&  \cite{Aidala:2018ond} \\ \hline 
          U+U & PHENIX & 192 GeV &$ 2.25-13$ & \cite{Acharya:2020xkf} \\ \hline 
          d+Au &PHENIX & 200 GeV & $2.25-11$  & \cite{Adler:2006bv} \\ \hline 
          Au+Au &PHENIX & 200 GeV &  $2.25-9.5$ & \cite{Adler:2006bv} \\ \hline 
          Au+Au &PHENIX & 200 GeV &  $5.5-17$  &\cite{Adare:2012wg}\\ \hline 
          Pb+Pb &ALICE &2.76 TeV &$1.25-18.5$ &\cite{Acharya:2018yhg} 
      \end{tabular}
      \ruledtabular
\end{table}



For this demonstration we adopt the functional form from  Eq.~\ref{Eq:empirical_fit_B} (empirical Fit B). The parameters are fixed using the simultaneous fit to the $p$+$p$ and $p$+A data to the following values: $ a= -1.24 $, $b=0.482$, $p_0=4.15$, $n=5.07$, and the composite parameter $R^\infty/A = 2.28$.   
The final fit parameter $R^\infty$ is determined individually for each data set using the data shuffling method. For each data set we vary the points many times within their systematic uncertainties, as discussed in Appendix~\ref{Appendix:shuffle}, and create an ensemble of $R^\infty$  and $\sigma_{R^\infty}$ values. The mean of the $R^\infty$ ensemble is used as the measurement of $R^\infty$ for ${\eta/\pi^0}$ and the standard deviation is quoted as the systematic uncertainty. The mean of the $\sigma_{R^\infty}$ ensemble is quoted as the statistical uncertainty. 

\begin{figure}
\includegraphics[width=2.8in]{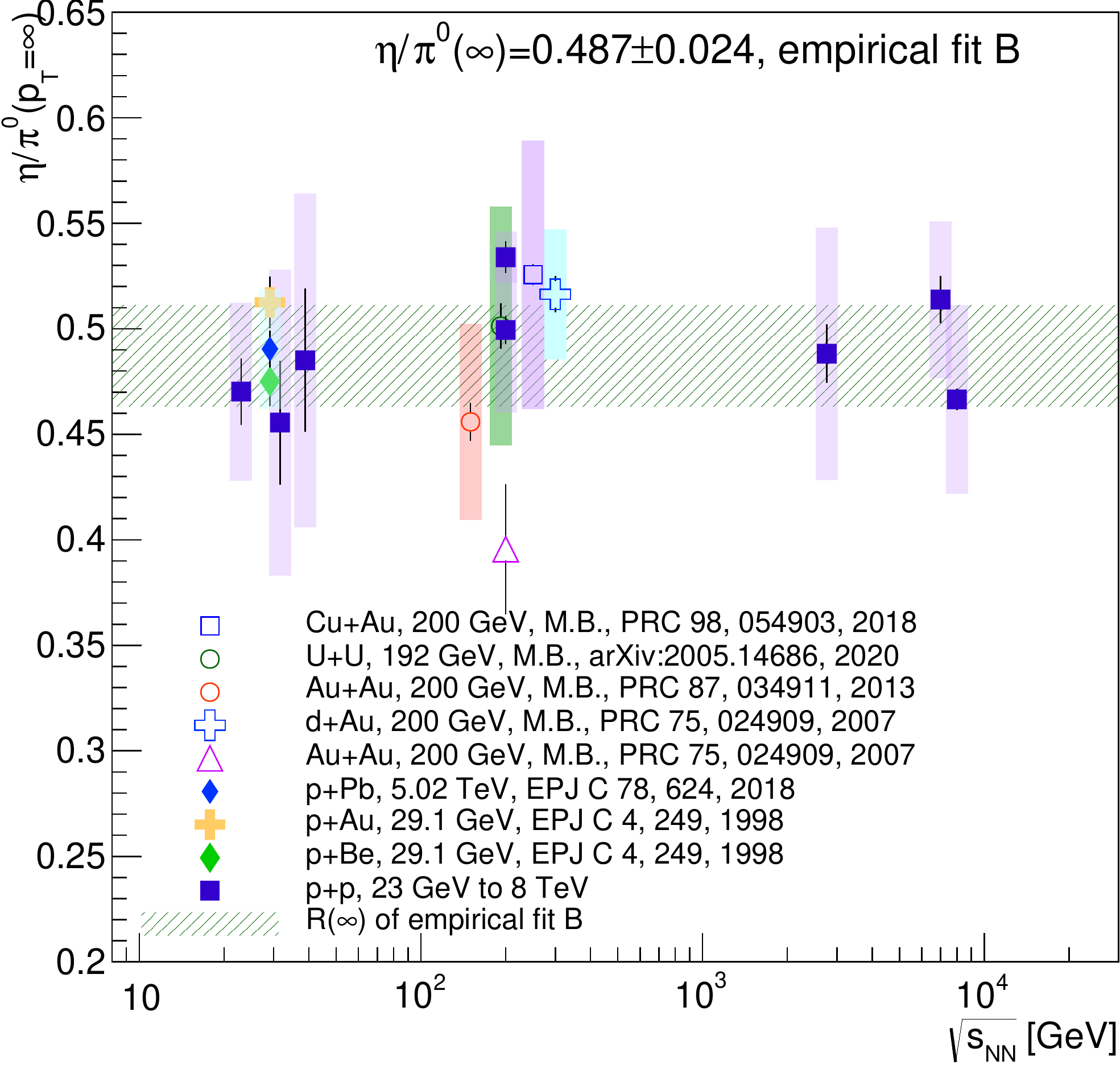}
\caption{Values of $R^\infty = \eta/\pi^0(p_T \rightarrow \infty)$ as a function of $\sqrt{s_{NN}}$ for the minimum bias $p$+$p$, $p$+A and A+B data sets. Statistical errors are shown as bars, systematic uncertainties as bands. Also shown is a band representing  $0.487\pm 0.024$, the result of the empirical fit B to the combined $p$+$p$ and $p$+A data. Note that the A+B data at 200 GeV are offset in $\sqrt{s_{NN}}$ to avoid overlap of data sets.}\label{Fig:Rinfty_vs_energy}
\end{figure}

Fig.~\ref{Fig:Rinfty_vs_energy} shows the results as a function of the nucleon-nucleon center of mass energy $\sqrt{s_{NN}}$ for the minimum bias data samples of all collision systems. Also shown on the figure is the $R^\infty$ value obtained from the combined fit to the $p$+$p$ and $p$+A data sets using method B. Within uncertainties all data sets are consistent with this value and there is no evidence for a $\sqrt{s_{NN}}$ dependence of $R^\infty$. 

For most publications of ${\eta/\pi^0}$ from heavy ion collisions, the data was also presented for centrality selected event classes. In order to include these in the comparison, we plot $R^\infty$ as a function of the number of produced particle $\eval{d N_{ch}/d{\eta}}_{\eta=0}$. The  $d{N_{ch}/{d\eta}}$ values used are summarized in Tab.~\ref{Tab:Nch}.

\begin{figure}
\includegraphics[height=2.8in]{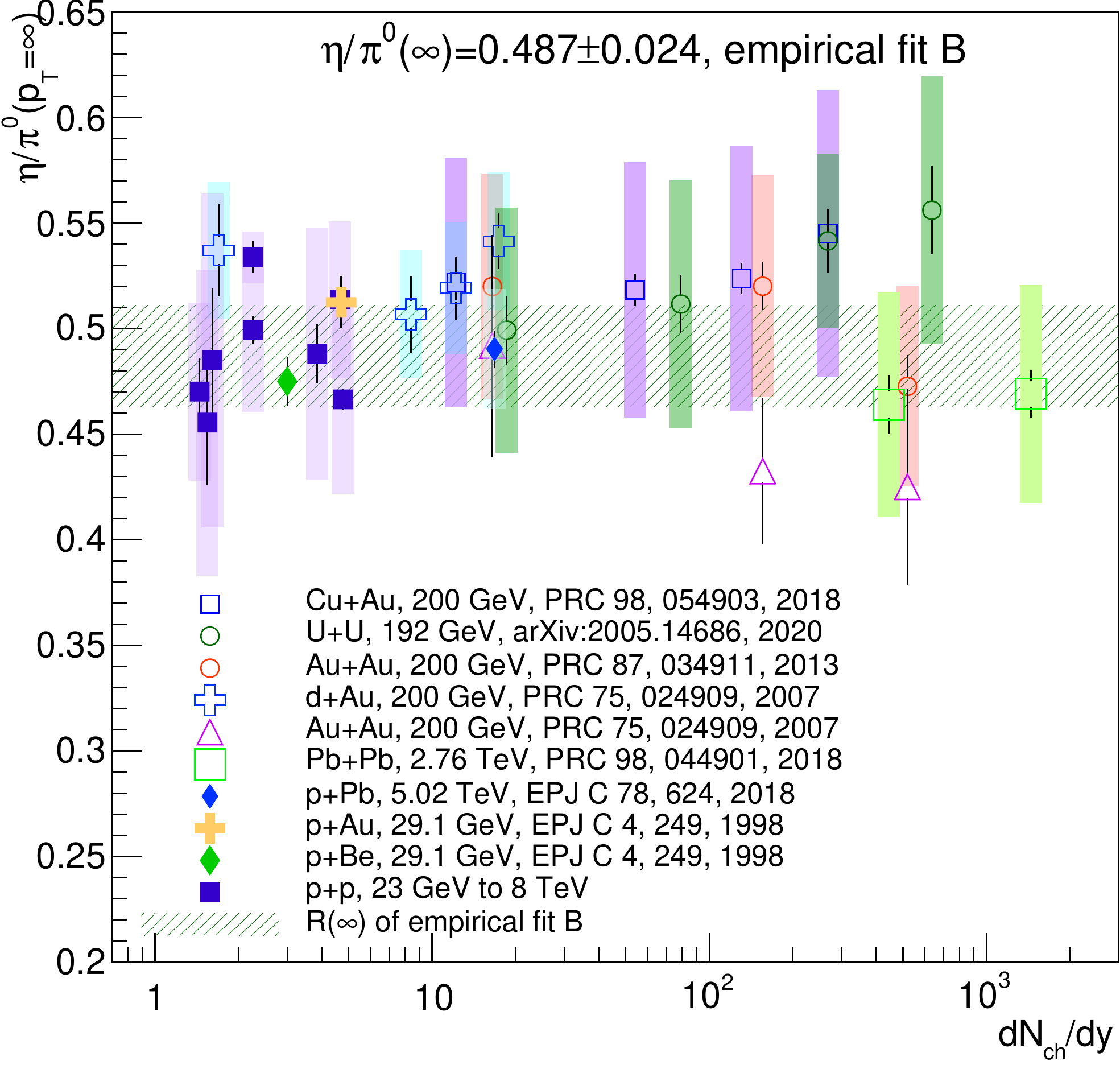}
\caption{Values of $R^\infty = \eta/\pi^0(p_T \rightarrow \infty)$ as a function of $dN_{ch}/d\eta$. The presentation is identical to Fig.~\ref{Fig:Rinfty_vs_energy}, however, for A+B collisions results from different centrality classes are shown rather than for the minimum bias sample. }
\label{Fig:Rinfty_vs_Nch}
\end{figure}

The results are given in Fig.~\ref{Fig:Rinfty_vs_Nch}. Again all values are consistent with a universal value within uncertainties. This analysis strongly suggest that $R^\infty$ does not depend on the collision systems, $\sqrt{s_{NN}}$, or the centrality of the collisions and that any apparent differences are likely due to  systematic effects specific to individual data sets.   

       
\begin{table}
\caption{Values for $\dd N_{ch}/\dd \eta$ at mid-rapidity for all collision systems and centrality selections used in this work. For p+p collisions, the numbers correspond to the inelastic $p$+$p$ cross section as given in \cite{Adam:2015gka}. For all other cases, whenever a reference is given, the values are taken directly for the publication. For PHENIX data we use data tabulated in \cite{Adare:2015bua}. The symbol * in the reference indicates that the value was extrapolated beyond what was tabulated. All minimum bias values (MB) that are marked by ** were calculated from the centrality selected data sets for the same system. For all data the uncertainties were calculated assuming that the values quoted in the reference are fully correlated. Reference \cite{Agakishiev:1998mw} does not give an uncertainty on the multiplicity value. }\label{Tab:Nch}
     \ruledtabular
      \begin{tabular}{c| c |c| c |  c} 
    System & $\sqrt{s_{NN}}$ & Centrality   &$\dd N_{ch}/\dd \eta$  & Ref.\\ \hline 
    p+p    & $\sqrt{s}$  & --  &$\alpha (\sqrt{s}/\text{GeV})^{2\delta}$ &   \cite{Adam:2015gka}  \\ \hline
    p+Au   &  29.1 GeV   & --          &   4.7            & \cite{Agakishiev:1998mw} \\ \hline
    p+Be   &  29.1 GeV   & --          &   3.0            & \cite{Agakishiev:1998mw} \\ \hline  
    p+Pb   &  5.02 TeV   & --          &   $16.8\pm0.7$   & \cite{ALICE:2012xs}    \\ \hline
    d+Au   &  200 GeV    & $0\%-20\%$  &   $17.4\pm1.2$   & \cite{Adare:2015bua} \\
           &             & $20\%-40\%$ &   $12.2\pm0.9$   & \cite{Adare:2015bua} \\
           &             & $40\%-60\%$ &   $8.4 \pm0.6$   & \cite{Adare:2015bua} \\
           &             & $60\%-88\%$ &   $1.7 \pm0.4$   & * \\
           &             & MB          &   $9.2\pm0.8$   & ** \\ \hline 
  Cu+Au    &  200 GeV    & $0\%-20\%$  & $268\pm20$      & \cite{Adare:2015bua} \\ 
           &             & $20\%-40\%$ & $131\pm10$      & \cite{Adare:2015bua} \\
           &             & $40\%-60\%$ & $54 \pm4$       & \cite{Adare:2015bua} \\
           &             & $60\%-93\%$ & $12.2 \pm1.5$   & *  \\
           &             & MB          & $102 \pm9$      & ** \\ \hline    
 Au+Au     &  200 GeV    & $0\%-20\%$  & $519\pm26$      & \cite{Adare:2015bua} \\
           &			 & $20\%-60\%$ & $156\pm11$      & \cite{Adare:2015bua} \\
	 	   &        	 & $60\%-92\%$ & $16.5\pm2$    & *  \\
		   &             & MB          & $186\pm11$      & ** \\\hline   
 U+U      &  192 GeV     & $0\%-20\%$  & $636\pm51$      &  \cite{Adare:2015bua}   \\
          &              & $20\%-40\%$ & $268\pm21$      &  \cite{Adare:2015bua}   \\
          &              & $40\%-60\%$ & $ 79\pm8$       &  *                      \\
          &              & $60\%-80\%$ & $ 18.6\pm3 $       &  *                      \\ 
          &              &  MB         & $ 234\pm35$        &  **     \\ \hline
 Pb+Pb    & 2.76 TeV     & $0\%-10\%$  & $1448\pm55$     &\cite{Aamodt:2010cz}\\ 
          &              & $20\%-50\%$ & $445.3\pm10$    &\cite{Aamodt:2010cz}\\  
      \end{tabular}
      \ruledtabular
\end{table}
\section{The effect of radial flow}\label{Sec:flow}

We have shown that $\eta/\pi^0$ can be described by one common function for all $p$+$p$ and $p$+A collisions over the measured $p_T$ range from 0.1 to 20 GeV/$c$. Furthermore, above $p_T$=5 GeV/$c$ the same function describes all data from heavy ion collisions. Whether this universal function also describes heavy ion data at lower $p_T$ can not be tested due to the absence of accurate experimental data. However, there are reasons to believe that this universality does not hold at low $p_T$. 

\begin{figure}
\centering
\includegraphics[width=2.8in]{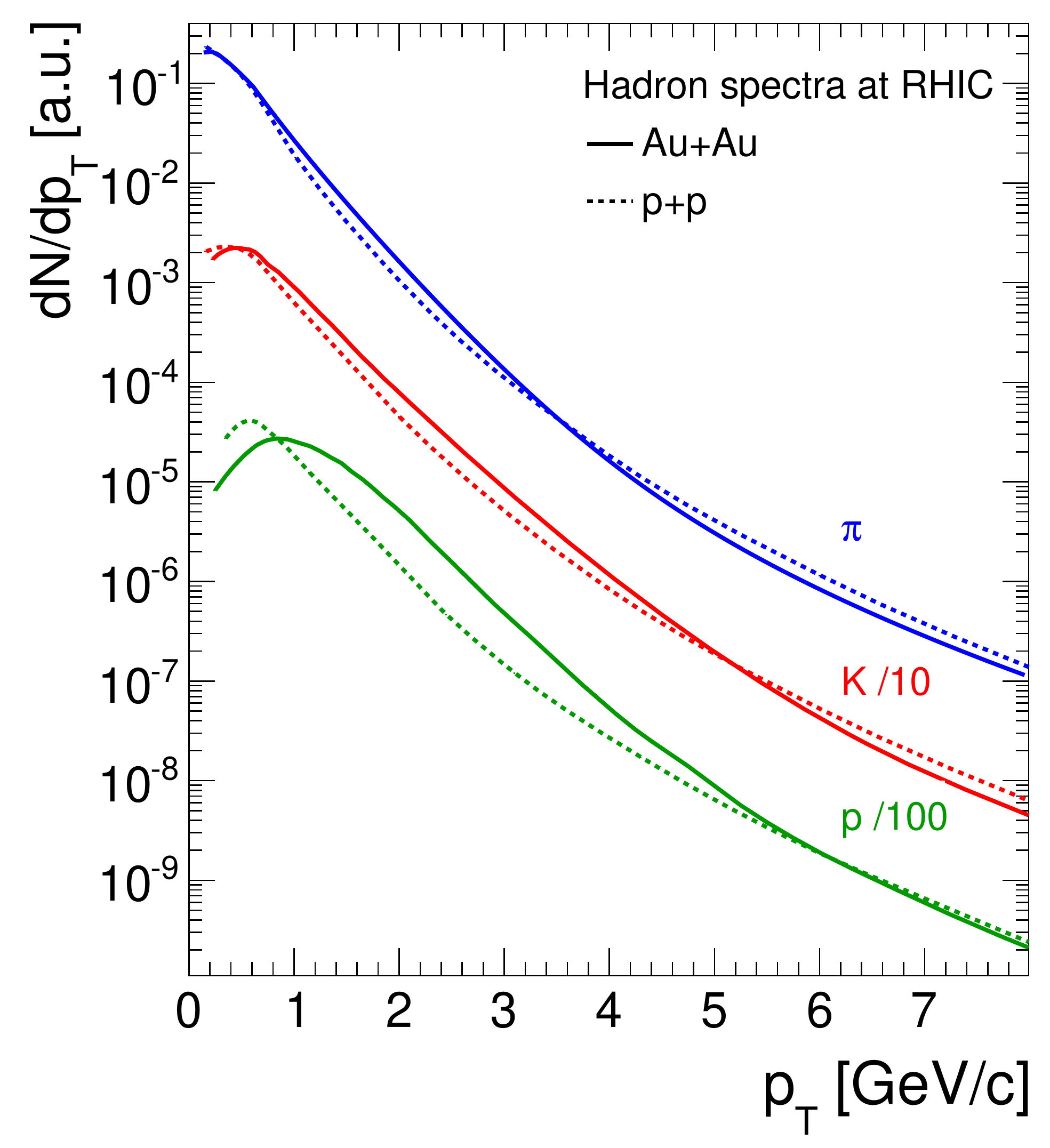}
\caption{Schematic comparison of $\pi, K, p$ spectra from Au+Au and p+p collisions at $\sqrt{s_{NN}}$=200 GeV. All spectra are approximately normalize to their rapidity density at mid rapidity. Different particle types are separated by factors of 10 for clarity.  }
\label{Fig:flow_cartoon}
\end{figure}

Evidence for strong collective motion of the bulk of the produced particles has been observed in all high energy heavy ion collision. 
This motion is consistent with a Hubble like hydrodynamic expansion of the collision volume, with a linear velocity profile in radial direction. In this velocity profile heavier particles gain more momentum than lighter ones. Radial flow effectively depletes the particle yields at low $p_T$ and enhances them in an intermediate $p_T$ range, which is determined by the mass of the particle. For $p_T$ much larger than the particle's mass radial flow becomes negligible. Fig.~\ref{Fig:flow_cartoon}  shows the effect schematically by comparing $\pi, K$, and $p$ spectra at RHIC energies. The spectra shown are roughly to scale and consistent with experimental data from 200 GeV Au+Au collisions. They are normalized per particle of the corresponding type at mid rapidity.  

Since the $\eta$ meson has about the same mass as the kaon, one would expect that in the momentum range from a few hundred MeV/$c$ to a few GeV/$c$ radial flow increases the yield of $\eta$ mesons significantly more than that of $\pi^0$. This in turn would increase the $\eta/\pi^0$ ratio in heavy ion collisions compared to that observed in $p$+$p$ and $p$+A collisions.   




To quantify the size of the modification due to radial flow we will use a double ratio  $R_{flow}$ defined as follows: 

\begin{equation}\label{Eq:R_flow}
R_{flow}\equiv \frac{\qty(\frac{\eta}{\pi^0})_{C_i}}{\qty(\frac{\eta}{\pi^0})_{p+p}} \approx \frac{\qty(\frac{K^\pm}{\pi^\pm})_{C_i}}{\qty(\frac{K^\pm}{\pi^\pm})_\text{p+p}}\equiv \frac{\qty(R_{AA}^{K^\pm})_{C_i}}{\qty(R_{AA}^{\pi^\pm})_{C_i}},
\end{equation}
where we take advantage of the fact the momentum boost from radial flow is mostly determined by the particle mass and that $m_{K^\pm}\approx m_{\eta}$. Also charged pions are used instead of neutral pions, since $\pi^\pm$ and kaons are typically measured simultaneously with the same detector systems and thus most systematic uncertainties on the  measurement cancel in the double ratio. The subscript $C_i$ refers to a specific collision system, energy and centrality selection.  

\begin{figure}[t]
 \centering
\includegraphics[height=2.8in]{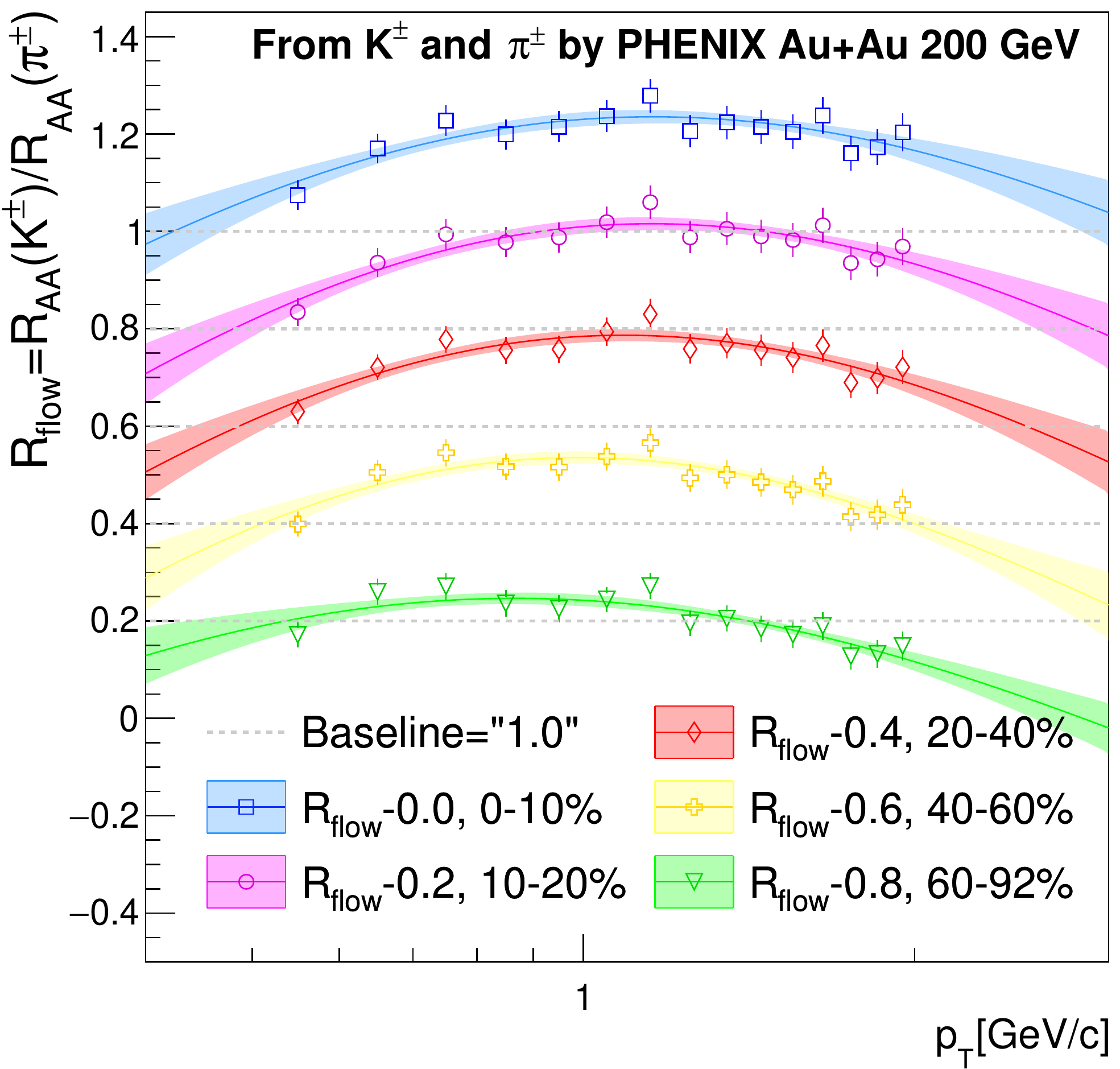}
\caption{Double ratio, obtained by $\qty(R_{AA}^K)_{C_i}/\qty(R_{AA}^\pi)_{C_i}$. }
\label{Fig:Rflow_PHENIX}
\end{figure}

Fig.~\ref{Fig:Rflow_PHENIX} presents $R_{flow}$ for different centrality classes of Au+Au collisions at 200 GeV. The values were calculated from data published by PHENIX \cite{Adare:2013esx}. The data cover the $p_T$ range from 0.5 to 2 GeV/$c$ and GPR is used to extrapolate somewhat beyond the measured range. According to this estimate the $\eta/\pi^0$ ratio is enhanced in central collisions in a $p_T$ region from 0.4 to 3 GeV/$c$ with a maximum of about 25\% near 1 GeV/c. The enhancement is reduced for more peripheral collisions and nearly vanishes for the 60-92\% selection. 

\begin{figure}
\centering
\includegraphics[width=2.8in]{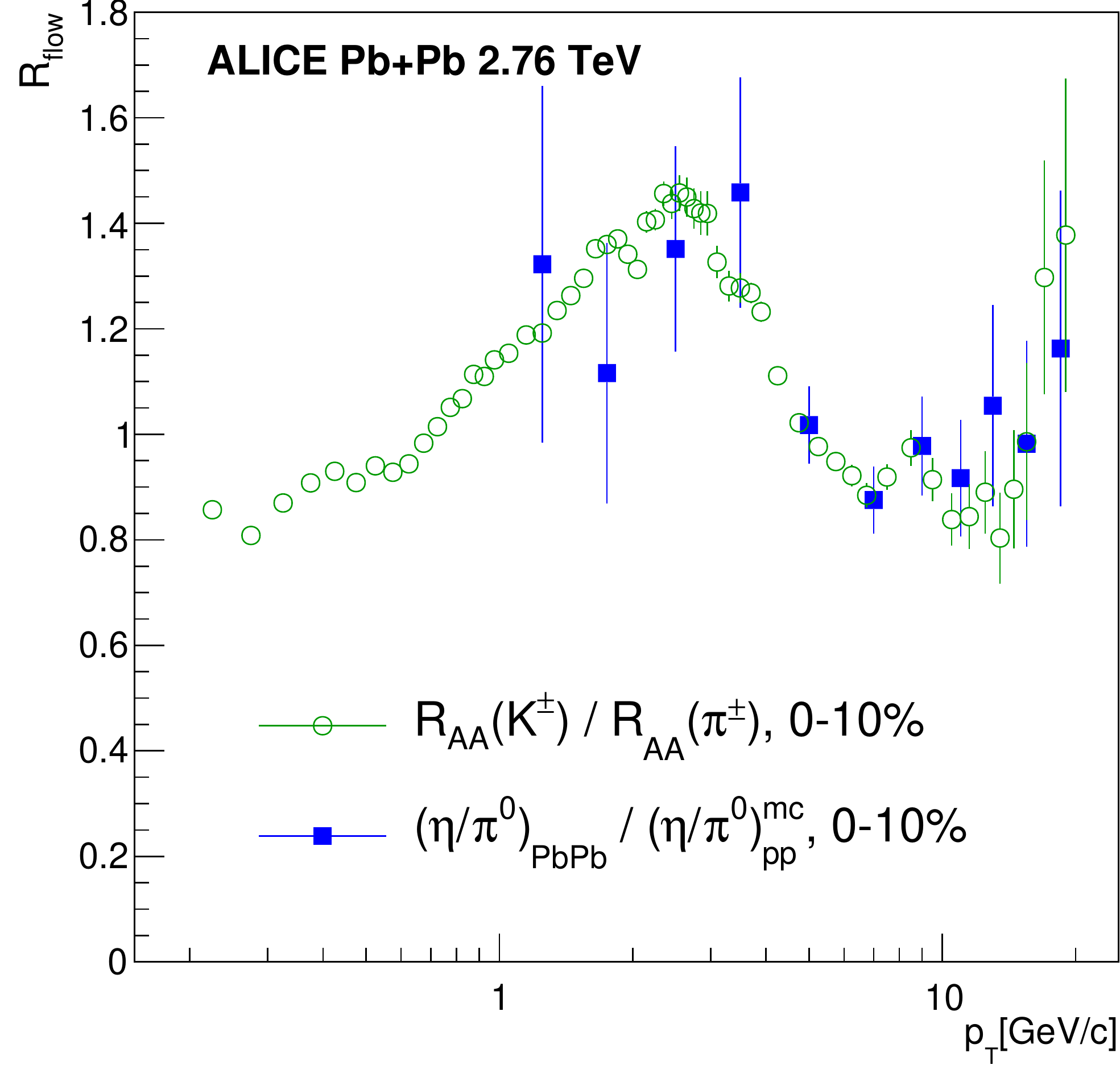}
\caption{Double ratio (the flow ratio) for $K^\pm/\pi^\pm$ and $\eta/\pi^0$ in Pb+Pb collisions at 2.76 TeV.
}
\label{Fig:Rflow-ALICE}
\end{figure}

In Fig.~\ref{Fig:Rflow-ALICE} we depict the estimate for $R_{flow}$ for Pb+Pb data at 2.76 TeV calculated from $K$ and $\pi^\pm$ data measured by ALICE \cite{Adam:2015kca,Acharya:2018yhg}. Shown are results for a 0-10\% centrality selections. 
Only statistical uncertainties are shown. 
The flow effect is significantly larger at LHC than at RHIC: the $p_T$ range affected is extended to 5-6 GeV/$c$, it reaches its maximum at higher $p_T$ around 3 GeV/$c$, and the maximum has increased to about 50\%. All indicates that radial flow effects increase with beam energy, which is consistent with a higher initial pressure and a longer lifetime of the system at the LHC compared to RHIC.  

ALICE also has published $\eta/\pi^0$ for Pb+Pb collisions at 2.76 TeV \cite{Acharya:2018yhg} down to 1 GeV/$c$, which can be used to verify the validity of the $R_{flow}$ estimate from $K/\pi$. For this we have divided Pb+Pb data by the universal $(\eta/\pi^0)^{mc}_{pp}$ from  Fig.~\ref{Fig:etapi_universal}. The result is also shown in Fig.~\ref{Fig:Rflow-ALICE}, error bars represent the combine uncertainty of  $(\eta/\pi^0)^{mc}_{pp}$ and the statistical uncertainty of $(\eta/\pi^0)_{PbPb}$. The ansatz that 
 $$\frac{(K^\pm/\pi^\pm)_{cent}}{(K^\pm/\pi^\pm)_{pp}}\approx\frac{(\eta/\pi^0)_{cent}}{ (\eta/\pi^0)_{pp} }$$ is consistent with the data.

To construct an $\eta/\pi^0$ ratio for a specific collision system and centrality selection we modify the universal shape $(\eta/\pi^0)^{mc}_{pp}$ determined from $p$+$p$ and $p$+A data (see Fig.~\ref{Fig:etapi_universal} from section~\ref{Sec:empirical_etapi}) with $R_{flow}$ for the selected heavy ion sample: 

\begin{equation}
\qty(\frac{\eta}{\pi^0})_{C_i} =  \qty(\frac{\eta}{\pi^0})_{pp}\times R_{flow}
\approx
\qty(\frac{\eta}{\pi^0})_{pp}\times  \frac{   \qty(\frac{K^\pm}{\pi^\pm})_{C_i}  }{ \qty(\frac{K^\pm}{\pi^\pm})_{pp} }.
\end{equation}

Since $R_{flow}$ may be available only in a limited $p_T$ region, for example from 0.4 to 2 GeV/$c$ in Fig.~\ref{Fig:Rflow_PHENIX}, we propose the following procedure that can be applied to any A+B collisions system if $\pi^\pm$ and $K$ data are available for the $p_T$ range affected by radial flow. In the first step we create pseudo data for $\eta/\pi^0$ by multiplying $R_{flow}$ point-by-point with the $(\eta/\pi^0)^{mc}_{pp}$ up to $p^{cut}_T$ where $R_{flow}(p_T)\approx 1$. This range extends to 1.4 or 4.5 GeV/c for Au+Au at 200 GeV at 60-92\% centrality and Pb+Pb at 2.76 TeV, respectively. To ensure that our flow estimate has the correct asymptotic behavior we add a second set of pseudo data with constant values of $\eta/\pi^0 = 0.487\pm0.024$. These are added either above 4 GeV/c where all data sets can be described by a constant (see section~\ref{Sec:universal_etapi}) or above $1.6 \times p_T^{cut}$, which ever is larger. The combine pseudo data are processed through a GPR to obtain a smooth curve. Finally, in order to account appropriately for the systematic uncertainties at high $p_T$ we merge the GPR describing the flow effect with $(\eta/\pi^0)^{mc}_{pp}$ above $p_T^{cut}$. The uncertainty band at low $p_T$ is also taken to be whichever is larger. 

\begin{figure}[t]
 \centering
\includegraphics[width=2.7in]{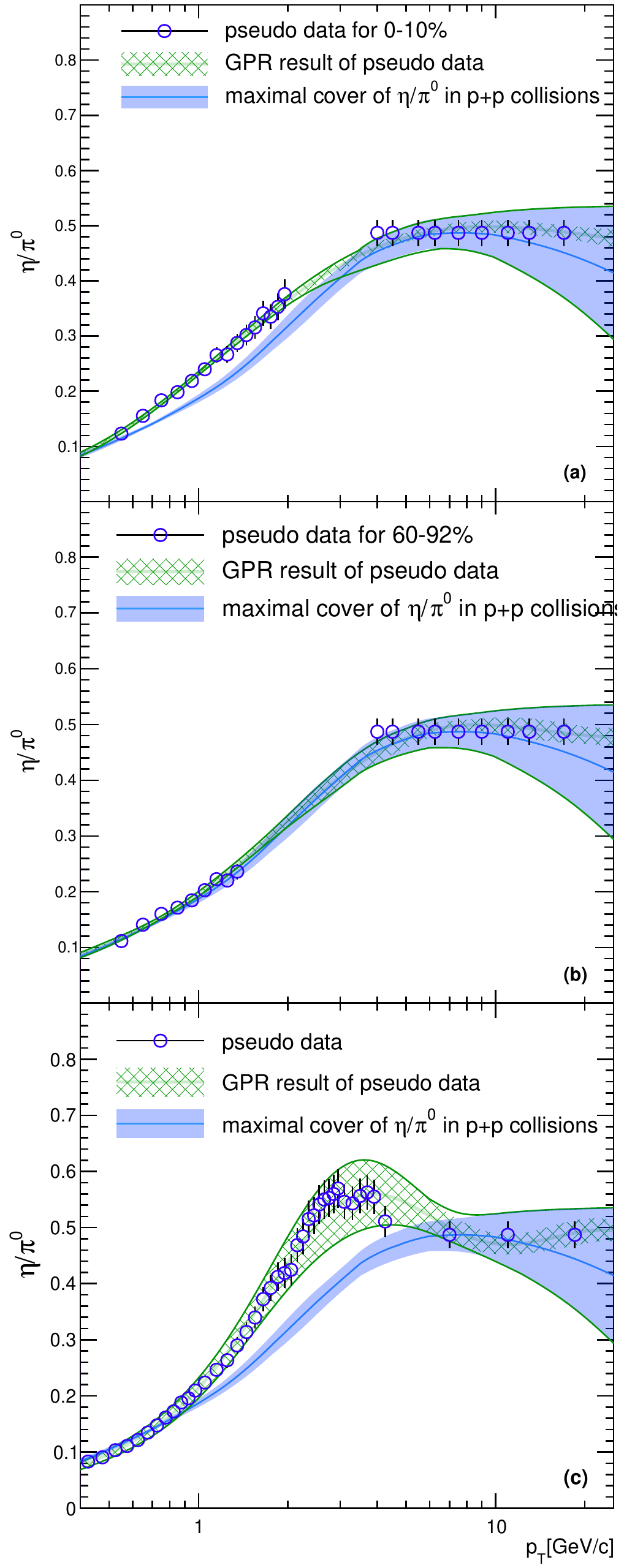}
\caption{Estimate of the effect of radial flow on $\eta/\pi^0$ for 0-20\%, 60-92\% Au+Au collisions at 200 GeV and 0-10\% Pb+Pb collisions at 2.76 TeV. Details are discussed in the text.  }\label{Fig:Construct-etapi-GPR}
\end{figure}
\begin{figure}[t]
 \centering
\includegraphics[width=2.7in]{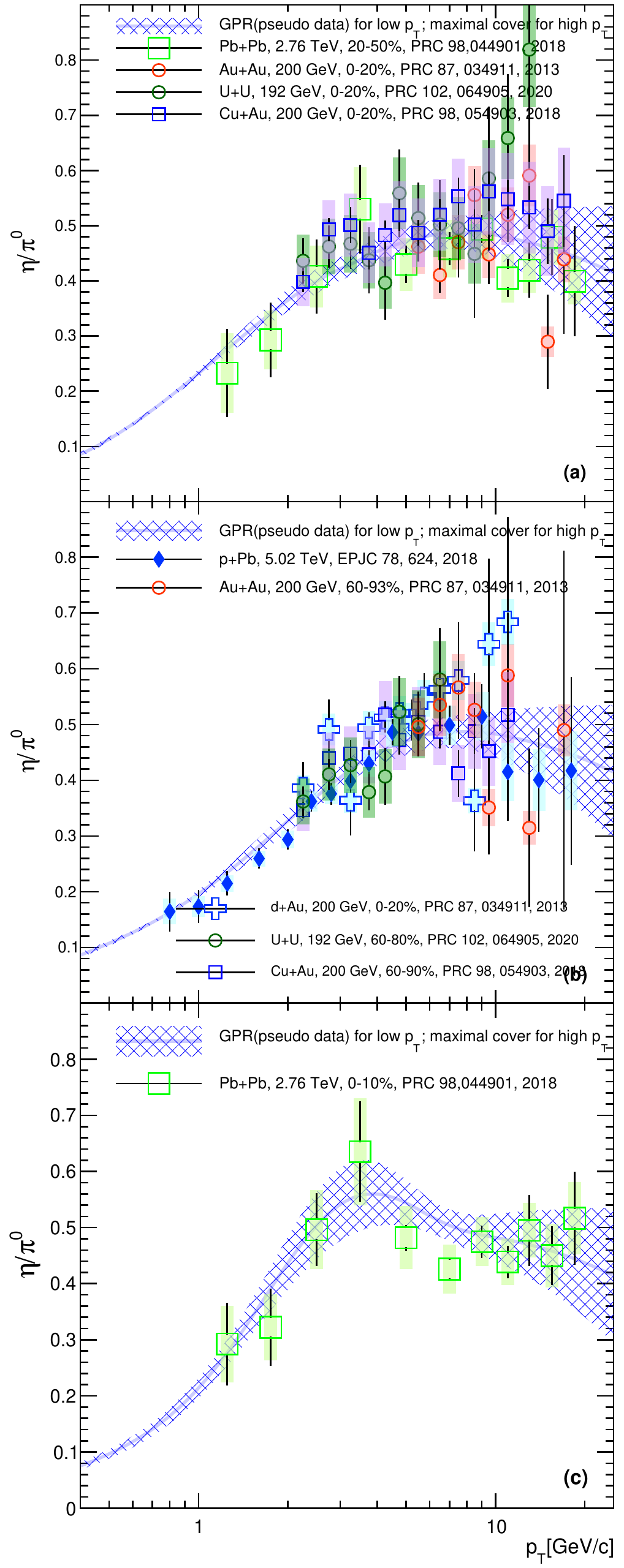}
\caption{Estimate of the effect of radial flow on $\eta/\pi^0$ for 0-20\%, 60-92\% Au+Au collisions at 200 GeV and 0-10\% Pb+Pb collisions at 2.76 TeV. Details are discussed in the text.  }\label{Fig:Construct-etapi-data}
\end{figure}

In Fig.~\ref{Fig:Construct-etapi-data} the construction $\eta/\pi^0$ is presented step by step for three examples: 0-20\%, 60-92\% Au+Au at 200 GeV and 0-10\% Pb+Pb at 2.76 TeV, in panels (a) to (c) respectively. The pseudo data generated are represented by points, which are then processed through a GPR resulting in the hashed green bands. They are contrasted with $(\eta/\pi^0)^{mc}_{pp}$, the blue band, and merged with it above $p_T^{cut}$ to create the final green envelope representing our $\eta/\pi^0$ estimates. As discussed above the largest flow effect is observed for central Pb+Pb collisions at the LHC (panel (c)). For central Au+Au collison at RHIC (panel (a)) a much smaller effect is observed, and finally peripheral collisions of the same system are consistent with no flow effect (panel (b)). 

These best estimates are compared to data in Fig.~\ref{Fig:Construct-etapi-GPR}. For the comparison we selected data sets with similar charged particle densities, so that despite the difference in collision system, centrality or $\sqrt{s_{NN}}$ matter was created under similar conditions and evolved the same way with time. In all three cases our best estimates are consistent with the $\eta/\pi^0$ data.

\section{Summary Discussion}\label{Sec:summary}


We find a universal $p_T$ dependence of $\eta/\pi^0$ for all $p$+$p$ and $p$+A collisions independent of the center of mass  energy from $\sqrt{s_{NN}}$=23 GeV  to 8 TeV. We note that like originally discovered in \cite{Agakishiev:1998mw}, below 3 GeV/$c$ the universal ratio is significantly below $m_T$ scaling extrapolations from higher $p_T$. 

That there is no $\sqrt{s_{NN}}$ dependence is surprising as the $p_T$ spectra of all particles vary strongly with $\sqrt{s_{NN}}$ and particle production from jet fragmentation becomes increasingly prevalent at higher energies. None-the-less there seems to be no impact on the relative yield at which $\eta$ and $\pi^0$ are produced. This may hint at a largely universal hadronisation process in which hadrons are always created under the same conditions, even if the underlying mechanism is considered different, for example bulk particle production or jet fragmentation.  

For heavy ion collisions, $\eta/\pi^0$ has the same universal behavior at high $p_T$, independent of collision species, collision energy, or collision centrality. For lower $p_T$ we find evidence for modifications of the relative particle yields due to radial flow. One might speculate that the same universal harmonization process is at work but that hadrons are produced in a moving reference frame. 

We have quantified the modification of the $\eta/\pi^0$ ratio due to radial flow using the double ratio $R_{AA}(K)/R_{AA}(\pi)$. This assumes that the change of the $p_T$ spectra depends entirely on the particle mass, but it does not make any assumptions about the similarity of $\eta$ and kaon spectra themselves. We note that our approach may overestimate the modification due to flow, since kaon production or generally strange quark production is enhanced in heavy ion collisions. In our estimate the modification increases with $\sqrt{s_{NN}}$. At 200 GeV at RHIC the maximum increase of $\eta/\pi^0$ is estimated to be 25\% around 1 GeV/$c$, in contrast at 2.76 TeV at the LHC the maximum increase is nearly 50\% and occurs at higher $p_T$ between 2 and 3 GeV/c. 

With our original motivation in mind, which was to reduce systematic uncertainties on the measurement of direct photons, we proposed a new methodology to create $\eta/\pi^0$ ratios. This method is more accurate than frequently used extrapolations to lower $p_T$ based on $m_T$ scaling, and does not suffer from the frequent lack of statistics for $\eta$ measurements. Our approach can be applied to all systems for which $K/\pi^\pm$ is measured in the $p_T$ range affected by radial flow. The method does not require actual measurements of $\eta$ production for a given system.  

We have tested this method for two specific collision systems. For Au+Au collisions at 200 GeV the deviations due to flow are found to be within $\pm$15\% of the minimum bias values. Even for central collisions the $\eta/\pi^0$ underlying the estimate of photon from hadron decays used in direct photon measurements \cite{Adare:2008ab, Adare:2014fwh,Adare:2018wgc} is above what we propose. As a consequence direct photon yields have been slightly under estimated, though the differences are within quoted systematic uncertainties.  
For central Pb+Pb collisions at 2.76 TeV the flow modifications are  larger and coincidentally bring $\eta/\pi^0$ much closer to the $m_T$ scaling assumption used in the measurement of direct photons published by ALICE \cite{Adam:2015lda}.

\begin{acknowledgments}
We acknowledge the support from the Office of Nuclear Physics in the Office of Science of the Department of Energy. 
\end{acknowledgments}

\appendix
\section{Gaussian Process Regression}\label{Appendix:GPR}

In this section we discuss the implementation of the  Gaussian Process Regression (GPR) used in our analysis. Full details about the GPR can be found in \cite{MIT_GPR}. We start with a selection of $N$ data points $x_i$, $y_i$, and $\sigma^2_i$. In our case this is typically, $N$ values of $\log10({p_T})$, $\eta/\pi^0(p_T)$ and its variance.   We use a  Square-Exponential (SE) kernel to describe the correlation between points, which is given by:

\begin{equation}
k_{SE}(x_i,x_j)=\sigma_p^2\exp(-\frac{(x_i-x_j)^2}{2l^2}).
\end{equation}

Here $\sigma_p$ gives the strength of the correlation between y values and $l$ is a length scale that determines the range in x over which y values are correlated.  

We introduce the vectors $X$ and $Y$, which have dimension $N$ and elements $x_i$ and $y_i$, i.e. the data. The correlations between y values is then defined by a covariance matrix $K_{xx}$ which has the elements

\begin{equation}
\qty(K_{xx})_{ij} = k_{SE}(x_i,x_j) + \delta_{ij} \sigma_i^2,
\end{equation}

with $\delta_{ii}=1$ and $\delta_{ij}=0$ for $i\neq j$. The term $ \delta_{ij} \sigma_i^2$ adds noise to the diagonal elements to account for the uncertainty on the measured y values. In order to determine $\sigma_p$ and $l$, we maximize the log likelihood function: 

\begin{align}
\log p(Y|\sigma_p,l)= & -\frac{n}{2}\log 2\pi
-\frac{1}{2}Y^T[K_{xx}]^{-1} Y \nnb\\
& -\frac{1}{2}\log \det(K_{xx}). \label{Eq:maxlike}
\end{align}

Once the parameters $\sigma_p$ and $l$ are set, we can predict y values for any given x value. For this we introduce a vectors $X^*$ and $Y^*$ of dimension R and elements $x^*_i$ for which we want to predict $y^*_i$, with R typically much larger than N. We introduce two more matrices, one of dimension $R\times N$ with elements $(K_{x^*x})_{ij}\equiv k_{SE}(x^*_i,x_j)$, and one of dimension $R\times R$ with elements $(K_{x^*x^*})_{ij}\equiv k_{SE}(x^*_i,x^*_j)$. The predicted values $Y^*$ and their covariance matrix $\text{Cov}(Y^*)$ are then calculated as follows:

\begin{equation}
Y^*=K_{x^*x}[K_{xx}]^{-1} Y,\label{Eq:Ypredict}
\end{equation}
\begin{equation}
\text{Cov}(Y^*)
=K_{x^*x^*}-K_{x^*x}[K_{xx}]^{-1}K_{x^*x}^T.
\end{equation}

The diagonal elements of $\text{Cov}(Y)$ give the variance of $Y^*$ due to the statistical uncertainty on the data $Y$. We refer to this as vector $S^*_{stat}$. We also consider the fit uncertainty on $\sigma_p$ and $l$. The variance can be calculated by the covariance matrix $M$ the fitting procedure provides through error propagation of Eq.~\ref{Eq:Ypredict}. 
:

\begin{align}
S_{fit}=&
(\partial_l y^*)^2 M_{ll} +2 \partial_ly^*\partial_{\sigma_p}y^* M_{l\sigma_p}\nnb\\
&+(\partial_{\sigma_p}y^*)^2 M_{\sigma_p\sigma_p},\quad\forall y^*\in Y^*\label{fit_GPR_err}
\end{align}

with $\partial_l$ and $\partial_{\sigma_f}$ being the partial derivatives of $Y^*$ with respect to $l$ and $\sigma_p$. 

In addition, we incorporate the systematic uncertainties using the data shuffling method discussed in Appendix~\ref{Appendix:shuffle}. We create a large  ensemble of different $\qty{Y^*_\lambda}$ for the same $X^*$ by varying each data set by a Gaussian random number $\epsilon\sim N(0,1)$  multiplying systematic uncertainties. The pointwise variance of ensambles $\qty{Y^*_\lambda}$, which we call  $S^*_{sys}$, is used as measure of the systematic uncertainty. 

In all figures that show results from the GPR the center line represents $Y^*$  and the vertical width of the band is $\sqrt{S_{stat}+S_{fit}+S^*_{Sys}}$, pointwise.


 
 
\section{Data-shuffling method}\label{Appendix:shuffle}
The \emph{data-shuffling method} is a Monte Carlo simulation approach that allows to estimate the effect of systematic uncertainties on the result of a fit of a function to data. To illustrate how the method works we first consider the case of one data set and assume that the systematic uncertainties are fully correlated. Here fully correlated means that the correlation matrix is $\rho_{ij}=1,\forall i,j$. Suppose each data point is described by a 4-tuple $(x_i,  y_i, \sigma^{stat}_i, \sigma^{sys}_i)$. One first defines a Gaussian random variable $\epsilon\sim N(0,1)$. In each simulation, one shifts each $y$ by a small quantity to $y\1_i=y_i+\sigma^{sys}_i\epsilon$ accordingly. Then in each simulation, one fits with these shifted data, and gets one fit result. This is repeated $L$ times, which generates $L$ sets of fit parameters. For each set of fit parameters one can devide the $x$ values into $R$ bins. Both $L$ and $R$ are usually large numbers. This results in a $L$-by-$R$ matrix of $y_{\lambda r}$ values. For a fixed $r$, the mean and standard deviation of $\qty{y_{\lambda r}:1\leq \lambda\leq L}$ are calculated. The standard deviation is assigned as systematic uncertainty of the fit for the given $r$. 

The method is expanded to multiple data sets by generating independent Gaussian random variables for each data set. In principle, more complex correlations of uncertainties for an individual data set can be decoded in $\rho_{ij}$, however, for the data at hand these correlations are not known and thus can not be implemented. 

One can choose as the final $y$ value for a given $r$ either the mean from data-shuffling, or  the fit result of the original data (i.e., the fit result when the Gaussian variables are zero). The difference between them is  usually negligible.

\bibliography{main.bib}

\begin{thebibliography}{30}%
\makeatletter
\providecommand \@ifxundefined [1]{%
 \@ifx{#1\undefined}
}%
\providecommand \@ifnum [1]{%
 \ifnum #1\expandafter \@firstoftwo
 \else \expandafter \@secondoftwo
 \fi
}%
\providecommand \@ifx [1]{%
 \ifx #1\expandafter \@firstoftwo
 \else \expandafter \@secondoftwo
 \fi
}%
\providecommand \natexlab [1]{#1}%
\providecommand \enquote  [1]{``#1''}%
\providecommand \bibnamefont  [1]{#1}%
\providecommand \bibfnamefont [1]{#1}%
\providecommand \citenamefont [1]{#1}%
\providecommand \href@noop [0]{\@secondoftwo}%
\providecommand \href [0]{\begingroup \@sanitize@url \@href}%
\providecommand \@href[1]{\@@startlink{#1}\@@href}%
\providecommand \@@href[1]{\endgroup#1\@@endlink}%
\providecommand \@sanitize@url [0]{\catcode `\\12\catcode `\$12\catcode
  `\&12\catcode `\#12\catcode `\^12\catcode `\_12\catcode `\%12\relax}%
\providecommand \@@startlink[1]{}%
\providecommand \@@endlink[0]{}%
\providecommand \url  [0]{\begingroup\@sanitize@url \@url }%
\providecommand \@url [1]{\endgroup\@href {#1}{\urlprefix }}%
\providecommand \urlprefix  [0]{URL }%
\providecommand \Eprint [0]{\href }%
\providecommand \doibase [0]{https://doi.org/}%
\providecommand \selectlanguage [0]{\@gobble}%
\providecommand \bibinfo  [0]{\@secondoftwo}%
\providecommand \bibfield  [0]{\@secondoftwo}%
\providecommand \translation [1]{[#1]}%
\providecommand \BibitemOpen [0]{}%
\providecommand \bibitemStop [0]{}%
\providecommand \bibitemNoStop [0]{.\EOS\space}%
\providecommand \EOS [0]{\spacefactor3000\relax}%
\providecommand \BibitemShut  [1]{\csname bibitem#1\endcsname}%
\let\auto@bib@innerbib\@empty
\bibitem [{\citenamefont {Shuryak}(1978)}]{Shuryak}%
  \BibitemOpen
  \bibfield  {author} {\bibinfo {author} {\bibfnamefont {E.~V.}\ \bibnamefont
  {Shuryak}},\ }\bibfield  {title} {\bibinfo {title} {{Quark-Gluon Plasma and
  Hadronic Production of Leptons, Photons and Psions}},\ }\href
  {https://doi.org/10.1016/0370-2693(78)90370-2} {\bibfield  {journal}
  {\bibinfo  {journal} {Sov. J. Nucl. Phys.}\ }\textbf {\bibinfo {volume}
  {28}},\ \bibinfo {pages} {408} (\bibinfo {year} {1978})}\BibitemShut
  {NoStop}%
\bibitem [{\citenamefont {Adare}\ \emph
  {et~al.}(2010{\natexlab{a}})\citenamefont {Adare} \emph
  {et~al.}}]{Adare:2008ab}%
  \BibitemOpen
  \bibfield  {author} {\bibinfo {author} {\bibfnamefont {A.}~\bibnamefont
  {Adare}} \emph {et~al.} (\bibinfo {collaboration} {PHENIX}),\ }\bibfield
  {title} {\bibinfo {title} {{Enhanced production of direct photons in Au+Au
  collisions at $\sqrt{s_{NN}}=200$ GeV and implications for the initial
  temperature}},\ }\href {https://doi.org/10.1103/PhysRevLett.104.132301}
  {\bibfield  {journal} {\bibinfo  {journal} {Phys. Rev. Lett.}\ }\textbf
  {\bibinfo {volume} {104}},\ \bibinfo {pages} {132301} (\bibinfo {year}
  {2010}{\natexlab{a}})},\ \Eprint {https://arxiv.org/abs/0804.4168}
  {arXiv:0804.4168 [nucl-ex]} \BibitemShut {NoStop}%
\bibitem [{\citenamefont {Adare}\ \emph {et~al.}(2015)\citenamefont {Adare}
  \emph {et~al.}}]{Adare:2014fwh}%
  \BibitemOpen
  \bibfield  {author} {\bibinfo {author} {\bibfnamefont {A.}~\bibnamefont
  {Adare}} \emph {et~al.} (\bibinfo {collaboration} {PHENIX}),\ }\bibfield
  {title} {\bibinfo {title} {{Centrality dependence of low-momentum
  direct-photon production in Au$+$Au collisions at $\sqrt{s_{_{NN}}}=200$
  GeV}},\ }\href {https://doi.org/10.1103/PhysRevC.91.064904} {\bibfield
  {journal} {\bibinfo  {journal} {Phys. Rev. C}\ }\textbf {\bibinfo {volume}
  {91}},\ \bibinfo {pages} {064904} (\bibinfo {year} {2015})},\ \Eprint
  {https://arxiv.org/abs/1405.3940} {arXiv:1405.3940 [nucl-ex]} \BibitemShut
  {NoStop}%
\bibitem [{\citenamefont {Adare}\ \emph {et~al.}(2019)\citenamefont {Adare}
  \emph {et~al.}}]{Adare:2018wgc}%
  \BibitemOpen
  \bibfield  {author} {\bibinfo {author} {\bibfnamefont {A.}~\bibnamefont
  {Adare}} \emph {et~al.} (\bibinfo {collaboration} {PHENIX}),\ }\bibfield
  {title} {\bibinfo {title} {{Beam Energy and Centrality Dependence of
  Direct-Photon Emission from Ultrarelativistic Heavy-Ion Collisions}},\ }\href
  {https://doi.org/10.1103/PhysRevLett.123.022301} {\bibfield  {journal}
  {\bibinfo  {journal} {Phys. Rev. Lett.}\ }\textbf {\bibinfo {volume} {123}},\
  \bibinfo {pages} {022301} (\bibinfo {year} {2019})},\ \Eprint
  {https://arxiv.org/abs/1805.04084} {arXiv:1805.04084 [hep-ex]} \BibitemShut
  {NoStop}%
\bibitem [{\citenamefont {Adam}\ \emph
  {et~al.}(2016{\natexlab{a}})\citenamefont {Adam} \emph
  {et~al.}}]{Adam:2015lda}%
  \BibitemOpen
  \bibfield  {author} {\bibinfo {author} {\bibfnamefont {J.}~\bibnamefont
  {Adam}} \emph {et~al.} (\bibinfo {collaboration} {ALICE}),\ }\bibfield
  {title} {\bibinfo {title} {{Direct photon production in Pb-Pb collisions at
  $\sqrt{s_{NN}} =$ 2.76 TeV}},\ }\href
  {https://doi.org/10.1016/j.physletb.2016.01.020} {\bibfield  {journal}
  {\bibinfo  {journal} {Phys. Lett. B}\ }\textbf {\bibinfo {volume} {754}},\
  \bibinfo {pages} {235} (\bibinfo {year} {2016}{\natexlab{a}})},\ \Eprint
  {https://arxiv.org/abs/1509.07324} {arXiv:1509.07324 [nucl-ex]} \BibitemShut
  {NoStop}%
\bibitem [{\citenamefont {Agakichiev}\ \emph {et~al.}(1998)\citenamefont
  {Agakichiev} \emph {et~al.}}]{Agakishiev:1998mw}%
  \BibitemOpen
  \bibfield  {author} {\bibinfo {author} {\bibfnamefont {G.}~\bibnamefont
  {Agakichiev}} \emph {et~al.},\ }\bibfield  {title} {\bibinfo {title}
  {{Neutral meson production in p Be and p Au collisions at 450-GeV beam
  energy}},\ }\href {https://doi.org/10.1007/s100529800804} {\bibfield
  {journal} {\bibinfo  {journal} {Eur. Phys. J. C}\ }\textbf {\bibinfo {volume}
  {4}},\ \bibinfo {pages} {249} (\bibinfo {year} {1998})}\BibitemShut {NoStop}%
\bibitem [{\citenamefont {Altenk\"amper}\ \emph {et~al.}(2017)\citenamefont
  {Altenk\"amper}, \citenamefont {Bock}, \citenamefont {Loizides},\ and\
  \citenamefont {Schmidt}}]{Altenkamper:2017qot}%
  \BibitemOpen
  \bibfield  {author} {\bibinfo {author} {\bibfnamefont {L.}~\bibnamefont
  {Altenk\"amper}}, \bibinfo {author} {\bibfnamefont {F.}~\bibnamefont {Bock}},
  \bibinfo {author} {\bibfnamefont {C.}~\bibnamefont {Loizides}},\ and\
  \bibinfo {author} {\bibfnamefont {N.}~\bibnamefont {Schmidt}},\ }\bibfield
  {title} {\bibinfo {title} {{Applicability of transverse mass scaling in
  hadronic collisions at energies available at the CERN Large Hadron
  Collider}},\ }\href {https://doi.org/10.1103/PhysRevC.96.064907} {\bibfield
  {journal} {\bibinfo  {journal} {Phys. Rev. C}\ }\textbf {\bibinfo {volume}
  {96}},\ \bibinfo {pages} {064907} (\bibinfo {year} {2017})},\ \Eprint
  {https://arxiv.org/abs/1710.01933} {arXiv:1710.01933 [hep-ph]} \BibitemShut
  {NoStop}%
\bibitem [{\citenamefont {Adare}\ \emph
  {et~al.}(2010{\natexlab{b}})\citenamefont {Adare} \emph {et~al.}}]{ppg088}%
  \BibitemOpen
  \bibfield  {author} {\bibinfo {author} {\bibfnamefont {A.}~\bibnamefont
  {Adare}} \emph {et~al.} (\bibinfo {collaboration} {PHENIX}),\ }\bibfield
  {title} {\bibinfo {title} {{ Detailed measurement of the $e^+ e^-$ pair
  continuum in $p+p$ and Au+Au collisions at $\sqrt{s_{NN}} = 200$ GeV and
  implications for direct photon production}},\ }\href
  {https://doi.org/10.1103/PhysRevC.81.034911} {\bibfield  {journal} {\bibinfo
  {journal} {Phys. Rev.}\ }\textbf {\bibinfo {volume} {C81}},\ \bibinfo {pages}
  {034911} (\bibinfo {year} {2010}{\natexlab{b}})},\ \Eprint
  {https://arxiv.org/abs/0912.0244} {arXiv:0912.0244 [nucl-ex]} \BibitemShut
  {NoStop}%
\bibitem [{\citenamefont {Hagedorn}(1965)}]{Hagedorn:1965st}%
  \BibitemOpen
  \bibfield  {author} {\bibinfo {author} {\bibfnamefont {R.}~\bibnamefont
  {Hagedorn}},\ }\bibfield  {title} {\bibinfo {title} {{Statistical
  thermodynamics of strong interactions at high-energies}},\ }\href@noop {}
  {\bibfield  {journal} {\bibinfo  {journal} {Nuovo Cim. Suppl.}\ }\textbf
  {\bibinfo {volume} {3}},\ \bibinfo {pages} {147} (\bibinfo {year}
  {1965})}\BibitemShut {NoStop}%
\bibitem [{\citenamefont {Acharya}\ \emph
  {et~al.}(2018{\natexlab{a}})\citenamefont {Acharya} \emph
  {et~al.}}]{Acharya:2017tlv}%
  \BibitemOpen
  \bibfield  {author} {\bibinfo {author} {\bibfnamefont {S.}~\bibnamefont
  {Acharya}} \emph {et~al.} (\bibinfo {collaboration} {ALICE}),\ }\bibfield
  {title} {\bibinfo {title} {{$\pi ^{0}$ and $\eta $ meson production in
  proton-proton collisions at $\sqrt{s}=8$ TeV}},\ }\href
  {https://doi.org/10.1140/epjc/s10052-018-5612-8} {\bibfield  {journal}
  {\bibinfo  {journal} {Eur. Phys. J. C}\ }\textbf {\bibinfo {volume} {78}},\
  \bibinfo {pages} {263} (\bibinfo {year} {2018}{\natexlab{a}})},\ \Eprint
  {https://arxiv.org/abs/1708.08745} {arXiv:1708.08745 [hep-ex]} \BibitemShut
  {NoStop}%
\bibitem [{\citenamefont {Abelev}\ \emph {et~al.}(2012)\citenamefont {Abelev}
  \emph {et~al.}}]{Abelev:2012cn}%
  \BibitemOpen
  \bibfield  {author} {\bibinfo {author} {\bibfnamefont {B.}~\bibnamefont
  {Abelev}} \emph {et~al.} (\bibinfo {collaboration} {ALICE}),\ }\bibfield
  {title} {\bibinfo {title} {{Neutral pion and $\eta$ meson production in
  proton-proton collisions at $\sqrt{s}=0.9$ TeV and $\sqrt{s}=7$ TeV}},\
  }\href {https://doi.org/10.1016/j.physletb.2012.09.015} {\bibfield  {journal}
  {\bibinfo  {journal} {Phys. Lett. B}\ }\textbf {\bibinfo {volume} {717}},\
  \bibinfo {pages} {162} (\bibinfo {year} {2012})},\ \Eprint
  {https://arxiv.org/abs/1205.5724} {arXiv:1205.5724 [hep-ex]} \BibitemShut
  {NoStop}%
\bibitem [{\citenamefont {Adler}\ \emph {et~al.}(2007)\citenamefont {Adler}
  \emph {et~al.}}]{Adler:2006bv}%
  \BibitemOpen
  \bibfield  {author} {\bibinfo {author} {\bibfnamefont {S.}~\bibnamefont
  {Adler}} \emph {et~al.} (\bibinfo {collaboration} {PHENIX}),\ }\bibfield
  {title} {\bibinfo {title} {{High transverse momentum $\eta$ meson production
  in $p^+ p$, $d^+$ Au and Au+Au collisions at $\sqrt{s_{NN} }$ = 200-GeV}},\
  }\href {https://doi.org/10.1103/PhysRevC.75.024909} {\bibfield  {journal}
  {\bibinfo  {journal} {Phys. Rev. C}\ }\textbf {\bibinfo {volume} {75}},\
  \bibinfo {pages} {024909} (\bibinfo {year} {2007})},\ \Eprint
  {https://arxiv.org/abs/nucl-ex/0611006} {arXiv:nucl-ex/0611006} \BibitemShut
  {NoStop}%
\bibitem [{\citenamefont {Acharya}\ \emph {et~al.}(2017)\citenamefont {Acharya}
  \emph {et~al.}}]{Acharya:2017hyu}%
  \BibitemOpen
  \bibfield  {author} {\bibinfo {author} {\bibfnamefont {S.}~\bibnamefont
  {Acharya}} \emph {et~al.} (\bibinfo {collaboration} {ALICE}),\ }\bibfield
  {title} {\bibinfo {title} {{Production of ${\pi ^0}$ and $\eta $ mesons up to
  high transverse momentum in pp collisions at 2.76 TeV}},\ }\href
  {https://doi.org/10.1140/epjc/s10052-017-4890-x} {\bibfield  {journal}
  {\bibinfo  {journal} {Eur. Phys. J. C}\ }\textbf {\bibinfo {volume} {77}},\
  \bibinfo {pages} {339} (\bibinfo {year} {2017})},\ \Eprint
  {https://arxiv.org/abs/1702.00917} {arXiv:1702.00917 [hep-ex]} \BibitemShut
  {NoStop}%
\bibitem [{\citenamefont {Adare}\ \emph {et~al.}(2011)\citenamefont {Adare}
  \emph {et~al.}}]{Adare:2010cy}%
  \BibitemOpen
  \bibfield  {author} {\bibinfo {author} {\bibfnamefont {A.}~\bibnamefont
  {Adare}} \emph {et~al.} (\bibinfo {collaboration} {PHENIX}),\ }\bibfield
  {title} {\bibinfo {title} {{Cross section and double helicity asymmetry for
  $\eta$ mesons and their comparison to neutral pion production in $p+p$
  collisions at $\sqrt{s}$=200 GeV}},\ }\href
  {https://doi.org/10.1103/PhysRevD.83.032001} {\bibfield  {journal} {\bibinfo
  {journal} {Phys. Rev. D}\ }\textbf {\bibinfo {volume} {83}},\ \bibinfo
  {pages} {032001} (\bibinfo {year} {2011})},\ \Eprint
  {https://arxiv.org/abs/1009.6224} {arXiv:1009.6224 [hep-ex]} \BibitemShut
  {NoStop}%
\bibitem [{\citenamefont {Acharya}\ \emph
  {et~al.}(2018{\natexlab{b}})\citenamefont {Acharya} \emph
  {et~al.}}]{Acharya:2018hzf}%
  \BibitemOpen
  \bibfield  {author} {\bibinfo {author} {\bibfnamefont {S.}~\bibnamefont
  {Acharya}} \emph {et~al.} (\bibinfo {collaboration} {ALICE}),\ }\bibfield
  {title} {\bibinfo {title} {{Neutral pion and $\eta$ meson production in p-Pb
  collisions at $\sqrt{s_\mathrm{NN}} = 5.02$ TeV}},\ }\href
  {https://doi.org/10.1140/epjc/s10052-018-6013-8} {\bibfield  {journal}
  {\bibinfo  {journal} {Eur. Phys. J. C}\ }\textbf {\bibinfo {volume} {78}},\
  \bibinfo {pages} {624} (\bibinfo {year} {2018}{\natexlab{b}})},\ \Eprint
  {https://arxiv.org/abs/1801.07051} {arXiv:1801.07051 [nucl-ex]} \BibitemShut
  {NoStop}%
\bibitem [{\citenamefont {Adler}\ \emph {et~al.}(2003)\citenamefont {Adler}
  \emph {et~al.}}]{Adler:2003pb}%
  \BibitemOpen
  \bibfield  {author} {\bibinfo {author} {\bibfnamefont {S.}~\bibnamefont
  {Adler}} \emph {et~al.} (\bibinfo {collaboration} {PHENIX}),\ }\bibfield
  {title} {\bibinfo {title} {{Mid-rapidity neutral pion production in proton
  proton collisions at $\sqrt{s}$ = 200-GeV}},\ }\href
  {https://doi.org/10.1103/PhysRevLett.91.241803} {\bibfield  {journal}
  {\bibinfo  {journal} {Phys. Rev. Lett.}\ }\textbf {\bibinfo {volume} {91}},\
  \bibinfo {pages} {241803} (\bibinfo {year} {2003})},\ \Eprint
  {https://arxiv.org/abs/hep-ex/0304038} {arXiv:hep-ex/0304038} \BibitemShut
  {NoStop}%
\bibitem [{\citenamefont {Sjostrand}\ \emph {et~al.}(2001)\citenamefont
  {Sjostrand}, \citenamefont {Eden}, \citenamefont {Friberg}, \citenamefont
  {Lonnblad}, \citenamefont {Miu}, \citenamefont {Mrenna},\ and\ \citenamefont
  {Norrbin}}]{Sjostrand:2000wi}%
  \BibitemOpen
  \bibfield  {author} {\bibinfo {author} {\bibfnamefont {T.}~\bibnamefont
  {Sjostrand}}, \bibinfo {author} {\bibfnamefont {P.}~\bibnamefont {Eden}},
  \bibinfo {author} {\bibfnamefont {C.}~\bibnamefont {Friberg}}, \bibinfo
  {author} {\bibfnamefont {L.}~\bibnamefont {Lonnblad}}, \bibinfo {author}
  {\bibfnamefont {G.}~\bibnamefont {Miu}}, \bibinfo {author} {\bibfnamefont
  {S.}~\bibnamefont {Mrenna}},\ and\ \bibinfo {author} {\bibfnamefont
  {E.}~\bibnamefont {Norrbin}},\ }\bibfield  {title} {\bibinfo {title}
  {{High-energy physics event generation with PYTHIA 6.1}},\ }\href
  {https://doi.org/10.1016/S0010-4655(00)00236-8} {\bibfield  {journal}
  {\bibinfo  {journal} {Comput. Phys. Commun.}\ }\textbf {\bibinfo {volume}
  {135}},\ \bibinfo {pages} {238} (\bibinfo {year} {2001})},\ \Eprint
  {https://arxiv.org/abs/hep-ph/0010017} {arXiv:hep-ph/0010017} \BibitemShut
  {NoStop}%
\bibitem [{\citenamefont {Rasmussen}\ and\ \citenamefont
  {Williams}(2005)}]{MIT_GPR}%
  \BibitemOpen
  \bibfield  {author} {\bibinfo {author} {\bibfnamefont {C.}~\bibnamefont
  {Rasmussen}}\ and\ \bibinfo {author} {\bibfnamefont {C.}~\bibnamefont
  {Williams}},\ }\href@noop {} {\emph {\bibinfo {title} {{Gaussian Processes
  for Machine Learning}}}}\ (\bibinfo  {publisher} {The MIT Press},\ \bibinfo
  {year} {2005})\BibitemShut {NoStop}%
\bibitem [{\citenamefont {Bonesini}\ \emph {et~al.}(1989)\citenamefont
  {Bonesini} \emph {et~al.}}]{ref80}%
  \BibitemOpen
  \bibfield  {author} {\bibinfo {author} {\bibfnamefont {M.}~\bibnamefont
  {Bonesini}} \emph {et~al.} (\bibinfo {collaboration} {WA70 Collaboration}),\
  }\bibfield  {title} {\bibinfo {title} {High transverse momentum $\eta$
  production in $\pi^-$p, $\pi^+$p, and pp interactions at 280 gev/c},\ }\href
  {https://doi.org/10.1007/BF01557657} {\bibfield  {journal} {\bibinfo
  {journal} {Z. Phys. C}\ }\textbf {\bibinfo {volume} {42}},\ \bibinfo {pages}
  {527} (\bibinfo {year} {1989})}\BibitemShut {NoStop}%
\bibitem [{\citenamefont {Apanasevich}\ \emph {et~al.}(2003)\citenamefont
  {Apanasevich} \emph {et~al.}}]{Apanasevich:2002wt}%
  \BibitemOpen
  \bibfield  {author} {\bibinfo {author} {\bibfnamefont {L.}~\bibnamefont
  {Apanasevich}} \emph {et~al.} (\bibinfo {collaboration} {Fermilab E706}),\
  }\bibfield  {title} {\bibinfo {title} {{Production of $\pi^0$ and $\eta$
  mesons at large transverse momenta in $pp$ and $pBe$ Interactions at 530 and
  800 GeV/c.}},\ }\href {https://doi.org/10.1103/PhysRevD.68.052001} {\bibfield
   {journal} {\bibinfo  {journal} {Phys. Rev. D}\ }\textbf {\bibinfo {volume}
  {68}},\ \bibinfo {pages} {052001} (\bibinfo {year} {2003})},\ \Eprint
  {https://arxiv.org/abs/hep-ex/0204031} {arXiv:hep-ex/0204031} \BibitemShut
  {NoStop}%
\bibitem [{\citenamefont {Aidala}\ \emph {et~al.}(2018)\citenamefont {Aidala}
  \emph {et~al.}}]{Aidala:2018ond}%
  \BibitemOpen
  \bibfield  {author} {\bibinfo {author} {\bibfnamefont {C.}~\bibnamefont
  {Aidala}} \emph {et~al.} (\bibinfo {collaboration} {PHENIX}),\ }\bibfield
  {title} {\bibinfo {title} {{Production of $\pi^0$ and $\eta$ mesons in
  Cu$+$Au collisions at $\sqrt{s_{NN}}$=200 GeV}},\ }\href
  {https://doi.org/10.1103/PhysRevC.98.054903} {\bibfield  {journal} {\bibinfo
  {journal} {Phys. Rev. C}\ }\textbf {\bibinfo {volume} {98}},\ \bibinfo
  {pages} {054903} (\bibinfo {year} {2018})},\ \Eprint
  {https://arxiv.org/abs/1805.04389} {arXiv:1805.04389 [hep-ex]} \BibitemShut
  {NoStop}%
\bibitem [{\citenamefont {Acharya}\ \emph {et~al.}(2020)\citenamefont {Acharya}
  \emph {et~al.}}]{Acharya:2020xkf}%
  \BibitemOpen
  \bibfield  {author} {\bibinfo {author} {\bibfnamefont {U.}~\bibnamefont
  {Acharya}} \emph {et~al.} (\bibinfo {collaboration} {PHENIX}),\ }\bibfield
  {title} {\bibinfo {title} {{Production of $\pi^0$, $\eta$, and $K_S$ mesons
  in U$+$U collisions at $\sqrt{s_{_{NN}}}=192$ GeV}},\ }\href
  {https://doi.org/10.1103/PhysRevC.102.064905} {\bibfield  {journal} {\bibinfo
   {journal} {Phys. Rev. C}\ }\textbf {\bibinfo {volume} {102}},\ \bibinfo
  {pages} {064905} (\bibinfo {year} {2020})},\ \Eprint
  {https://arxiv.org/abs/2005.14686} {arXiv:2005.14686 [hep-ex]} \BibitemShut
  {NoStop}%
\bibitem [{\citenamefont {Adare}\ \emph
  {et~al.}(2013{\natexlab{a}})\citenamefont {Adare} \emph
  {et~al.}}]{Adare:2012wg}%
  \BibitemOpen
  \bibfield  {author} {\bibinfo {author} {\bibfnamefont {A.}~\bibnamefont
  {Adare}} \emph {et~al.} (\bibinfo {collaboration} {PHENIX}),\ }\bibfield
  {title} {\bibinfo {title} {{Neutral pion production with respect to
  centrality and reaction plane in Au$+$Au collisions at $\sqrt{s_{NN}}$=200
  GeV}},\ }\href {https://doi.org/10.1103/PhysRevC.87.034911} {\bibfield
  {journal} {\bibinfo  {journal} {Phys. Rev. C}\ }\textbf {\bibinfo {volume}
  {87}},\ \bibinfo {pages} {034911} (\bibinfo {year} {2013}{\natexlab{a}})},\
  \Eprint {https://arxiv.org/abs/1208.2254} {arXiv:1208.2254 [nucl-ex]}
  \BibitemShut {NoStop}%
\bibitem [{\citenamefont {Acharya}\ \emph
  {et~al.}(2018{\natexlab{c}})\citenamefont {Acharya} \emph
  {et~al.}}]{Acharya:2018yhg}%
  \BibitemOpen
  \bibfield  {author} {\bibinfo {author} {\bibfnamefont {S.}~\bibnamefont
  {Acharya}} \emph {et~al.} (\bibinfo {collaboration} {ALICE}),\ }\bibfield
  {title} {\bibinfo {title} {{Neutral pion and $\eta$ meson production at
  mid-rapidity in Pb-Pb collisions at $\sqrt{s_{NN}}$ = 2.76 TeV}},\ }\href
  {https://doi.org/10.1103/PhysRevC.98.044901} {\bibfield  {journal} {\bibinfo
  {journal} {Phys. Rev. C}\ }\textbf {\bibinfo {volume} {98}},\ \bibinfo
  {pages} {044901} (\bibinfo {year} {2018}{\natexlab{c}})},\ \Eprint
  {https://arxiv.org/abs/1803.05490} {arXiv:1803.05490 [nucl-ex]} \BibitemShut
  {NoStop}%
\bibitem [{\citenamefont {Adam}\ \emph {et~al.}(2017)\citenamefont {Adam} \emph
  {et~al.}}]{Adam:2015gka}%
  \BibitemOpen
  \bibfield  {author} {\bibinfo {author} {\bibfnamefont {J.}~\bibnamefont
  {Adam}} \emph {et~al.} (\bibinfo {collaboration} {ALICE}),\ }\bibfield
  {title} {\bibinfo {title} {{Charged-particle multiplicities in
  proton\textendash{}proton collisions at $\sqrt{s} = 0.9$ to 8 TeV}},\ }\href
  {https://doi.org/10.1140/epjc/s10052-016-4571-1} {\bibfield  {journal}
  {\bibinfo  {journal} {Eur. Phys. J. C}\ }\textbf {\bibinfo {volume} {77}},\
  \bibinfo {pages} {33} (\bibinfo {year} {2017})},\ \Eprint
  {https://arxiv.org/abs/1509.07541} {arXiv:1509.07541 [nucl-ex]} \BibitemShut
  {NoStop}%
\bibitem [{\citenamefont {Adare}\ \emph {et~al.}(2016)\citenamefont {Adare}
  \emph {et~al.}}]{Adare:2015bua}%
  \BibitemOpen
  \bibfield  {author} {\bibinfo {author} {\bibfnamefont {A.}~\bibnamefont
  {Adare}} \emph {et~al.} (\bibinfo {collaboration} {PHENIX}),\ }\bibfield
  {title} {\bibinfo {title} {{Transverse energy production and charged-particle
  multiplicity at midrapidity in various systems from $\sqrt{s_{NN}}=7.7$ to
  200 GeV}},\ }\href {https://doi.org/10.1103/PhysRevC.93.024901} {\bibfield
  {journal} {\bibinfo  {journal} {Phys. Rev. C}\ }\textbf {\bibinfo {volume}
  {93}},\ \bibinfo {pages} {024901} (\bibinfo {year} {2016})},\ \Eprint
  {https://arxiv.org/abs/1509.06727} {arXiv:1509.06727 [nucl-ex]} \BibitemShut
  {NoStop}%
\bibitem [{\citenamefont {Abelev}\ \emph {et~al.}(2013)\citenamefont {Abelev}
  \emph {et~al.}}]{ALICE:2012xs}%
  \BibitemOpen
  \bibfield  {author} {\bibinfo {author} {\bibfnamefont {B.}~\bibnamefont
  {Abelev}} \emph {et~al.} (\bibinfo {collaboration} {ALICE}),\ }\bibfield
  {title} {\bibinfo {title} {{Pseudorapidity density of charged particles in
  $p$ + Pb collisions at $\sqrt{s_{NN}}=5.02$ TeV}},\ }\href
  {https://doi.org/10.1103/PhysRevLett.110.032301} {\bibfield  {journal}
  {\bibinfo  {journal} {Phys. Rev. Lett.}\ }\textbf {\bibinfo {volume} {110}},\
  \bibinfo {pages} {032301} (\bibinfo {year} {2013})},\ \Eprint
  {https://arxiv.org/abs/1210.3615} {arXiv:1210.3615 [nucl-ex]} \BibitemShut
  {NoStop}%
\bibitem [{\citenamefont {Aamodt}\ \emph {et~al.}(2011)\citenamefont {Aamodt}
  \emph {et~al.}}]{Aamodt:2010cz}%
  \BibitemOpen
  \bibfield  {author} {\bibinfo {author} {\bibfnamefont {K.}~\bibnamefont
  {Aamodt}} \emph {et~al.} (\bibinfo {collaboration} {ALICE}),\ }\bibfield
  {title} {\bibinfo {title} {{Centrality dependence of the charged-particle
  multiplicity density at mid-rapidity in Pb-Pb collisions at
  $\sqrt{s_{NN}}=2.76$ TeV}},\ }\href
  {https://doi.org/10.1103/PhysRevLett.106.032301} {\bibfield  {journal}
  {\bibinfo  {journal} {Phys. Rev. Lett.}\ }\textbf {\bibinfo {volume} {106}},\
  \bibinfo {pages} {032301} (\bibinfo {year} {2011})},\ \Eprint
  {https://arxiv.org/abs/1012.1657} {arXiv:1012.1657 [nucl-ex]} \BibitemShut
  {NoStop}%
\bibitem [{\citenamefont {Adare}\ \emph
  {et~al.}(2013{\natexlab{b}})\citenamefont {Adare} \emph
  {et~al.}}]{Adare:2013esx}%
  \BibitemOpen
  \bibfield  {author} {\bibinfo {author} {\bibfnamefont {A.}~\bibnamefont
  {Adare}} \emph {et~al.} (\bibinfo {collaboration} {PHENIX}),\ }\bibfield
  {title} {\bibinfo {title} {{Spectra and ratios of identified particles in
  Au+Au and $d$+Au collisions at $\sqrt{s_{NN}}=200$ GeV}},\ }\href
  {https://doi.org/10.1103/PhysRevC.88.024906} {\bibfield  {journal} {\bibinfo
  {journal} {Phys. Rev. C}\ }\textbf {\bibinfo {volume} {88}},\ \bibinfo
  {pages} {024906} (\bibinfo {year} {2013}{\natexlab{b}})},\ \Eprint
  {https://arxiv.org/abs/1304.3410} {arXiv:1304.3410 [nucl-ex]} \BibitemShut
  {NoStop}%
\bibitem [{\citenamefont {Adam}\ \emph
  {et~al.}(2016{\natexlab{b}})\citenamefont {Adam} \emph
  {et~al.}}]{Adam:2015kca}%
  \BibitemOpen
  \bibfield  {author} {\bibinfo {author} {\bibfnamefont {J.}~\bibnamefont
  {Adam}} \emph {et~al.} (\bibinfo {collaboration} {ALICE}),\ }\bibfield
  {title} {\bibinfo {title} {{Centrality dependence of the nuclear modification
  factor of charged pions, kaons, and protons in Pb-Pb collisions at
  $\sqrt{s_{\rm NN}}=2.76$ TeV}},\ }\href
  {https://doi.org/10.1103/PhysRevC.93.034913} {\bibfield  {journal} {\bibinfo
  {journal} {Phys. Rev. C}\ }\textbf {\bibinfo {volume} {93}},\ \bibinfo
  {pages} {034913} (\bibinfo {year} {2016}{\natexlab{b}})},\ \Eprint
  {https://arxiv.org/abs/1506.07287} {arXiv:1506.07287 [nucl-ex]} \BibitemShut
  {NoStop}%
\end{thebibliography}%

\end{document}